\documentclass[aps,prb,twocolumn,notitlepage,superscriptaddress,am]{revtex4}%
\usepackage{graphicx}
\usepackage{amsmath}
\usepackage{amssymb}
\usepackage{color}
\usepackage{amsfonts}%
\providecommand{\U}[1]{\protect\rule{.1in}{.1in}}

\def\cm{cm$^{-1}$}

\def\stf{$\kappa$-STF$_x$}
\def\STF{$\kappa$-[(BEDT-STF)$_x$(BEDT-TTF)$_{1-x}$]$\rm _2\-Cu_2(CN)_3$}

\def\Cu{$\kappa$-(BEDT-TTF)$_2$Cu$_2$(CN)$_3$}

\begin{document}
\title{Rise and Fall of Landau's Quasiparticles While Approaching the Mott Transition}

\author{A. Pustogow\footnote{Present Address: Institute of Solid State Physics, Vienna University of Technology, 1040 Vienna, Austria.}}
\affiliation{1.~Physikalisches Institut, Universit\"{a}t Stuttgart, 70569 Stuttgart, Germany}
\affiliation{Department of Physics and Astronomy, UCLA, Los Angeles, California 90095, USA}
\author{Y. Saito}
\affiliation{1.~Physikalisches Institut, Universit\"{a}t Stuttgart, 70569 Stuttgart, Germany}
\affiliation{Department of Physics, Graduate School of Science, Hokkaido University, Sapporo 060-0810, Japan}
\author{A. L\"ohle}
\affiliation{1.~Physikalisches Institut, Universit\"{a}t Stuttgart, 70569 Stuttgart, Germany}
\author{M. Sanz Alonso}
\affiliation{1.~Physikalisches Institut, Universit\"{a}t Stuttgart, 70569 Stuttgart, Germany}
\author{A. Kawamoto}
\affiliation{Department of Physics, Graduate School of Science, Hokkaido University, Sapporo 060-0810, Japan}
\author{V. Dobrosavljevi\'c}
\affiliation{Department of Physics and National High Magnetic Field Laboratory, Florida State University, Tallahassee, Florida 32306, USA}
\author{M. Dressel}
\affiliation{1.~Physikalisches Institut, Universit\"{a}t Stuttgart, 70569 Stuttgart, Germany}\author{S. Fratini}
\affiliation{Institut N\'{e}el - CNRS and Universit\'{e} Grenoble Alpes, 38042 Grenoble Cedex 9, France}


\begin{abstract}
Landau suggested that the low-temperature properties of metals can be understood in terms of long-lived quasiparticles with all complex interactions included in Fermi-liquid parameters, such as the effective mass $m^{\star}$. Despite its wide applicability, electronic transport in bad or strange metals and unconventional superconductors is controversially discussed towards a possible collapse of the quasiparticle concept.
Here we explore the electrodynamic response of correlated metals at half filling for varying correlation strength upon approaching a Mott insulator. We reveal persistent Fermi-liquid behavior with pronounced quadratic dependences of the optical scattering rate on temperature and frequency, along with a puzzling elastic contribution to relaxation. The strong increase of the resistivity beyond the Ioffe-Regel-Mott limit is accompanied by a `displaced Drude peak' in the optical conductivity.
Our results, supported by a theoretical model for the optical response, demonstrate the emergence of a bad metal from resilient quasiparticles that are subject to dynamical localization and dissolve near the Mott transition.
\end{abstract}

\maketitle

\begin{figure}[t]
\centering
\includegraphics[width=1.0\columnwidth]{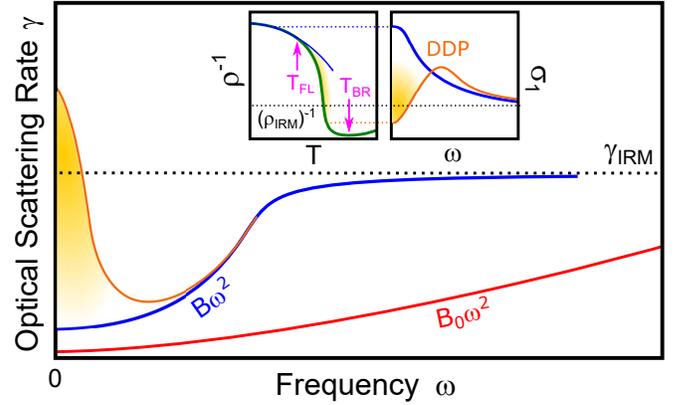}
\caption{\textbf{Scattering rate of correlated metals.}
\footnotesize{
In a Fermi liquid $\gamma(T,\omega)$ scales with $T^2$ and $\omega^2$ due to the increase in scattering phase space (Eq.~\ref{FL-Gurzhi}). In common metals, electron-electron scattering is weak (red) and other dissipation processes dominate.
The $\omega^2$ dependence prevails (blue) as electronic correlations yield $B\gg B_0$.
While in good metals $\gamma(T,\omega)$ saturates at the Ioffe-Regel-Mott (IRM) limit\cite{Gunnarsson2003,Hussey2004}, dynamical localization\cite{Fratini2016} can exceed this bound at low frequencies (orange). Insets: At $T<T_{\rm FL}$ the quasiparticle peak in the optical conductivity $\sigma_1$ occurs at $\omega=0$. At $T_{\rm FL}<T<T_{\rm BR}$, the resistivity ($\rho^{-1}$ shown to compare with $\sigma_1$) deviates from $\rho\propto T^2$ (blue) and increases beyond $\rho_{\rm IRM}$, which yields a drop of $\sigma_1$ at low frequencies, forming a `displaced Drude peak' (DDP) in a bad-metallic state.
}
}
\label{scattering-sketch}
\end{figure}

\textbf{Introduction} \quad
Conduction electrons in solids behave differently compared to free charges in vacuum. Since it is not possible to exhaustively model the interactions with all constituents of the crystal (nuclei and other electrons), Landau postulated quasiparticles (QP) with charge $e$ and spin $\frac{1}{2}$, which can be treated as nearly-free electrons but carry a renormalized mass $m^{\star}$ that incorporates all interaction effects\cite{Landau1956}.
In his Fermi-liquid picture, the conductivity of metals scales with the QP lifetime $\tau$, which increases asymptotically at low energy as the scattering phase space shrinks to zero\cite{Landau1956}. 
Electron-electron interaction involves a quadratic energy dependence of the scattering rate $\gamma=\tau^{-1}$ on both  temperature $T$ and frequency $\omega$,\cite{Gurzhi1959,Maslov2012,Berthod2013,Maslov2017} expressed as:
\begin{equation}
\gamma(T,\omega)=\gamma_{0}+B \left[ (p \pi k_{\rm B}/\hbar)^2 T^2+\omega^2 \right]\quad .
\label{FL-Gurzhi}
\end{equation}
Here $\gamma_{0}$ stems from residual scattering processes at zero energy (e.g. impurities), and $p$ is a numerical constant; the coefficient $B$ controls the overall rate of variation with energy and increases with the effective mass $m^{\star}$.
 In most metals with large electronic bandwidth $W$ the  behavior described by Eq. (\ref{FL-Gurzhi}) is not seen because of the small $m^{\star}$, so that the intrinsic contribution to scattering is negligible compared to other sources of dissipation (Fig.~\ref{scattering-sketch}).
Electronic correlations can strongly enhance the effective mass, ${m^{\star}}/{m_{b}}\gg 1$ (in a local Fermi liquid the QP weight scales as $Z\propto (m^*/m_b)^{-1}$, where $m_b$ is the band mass), making the energy-dependent terms of Eq.~\ref{FL-Gurzhi} the dominant contributions in the dc resistivity $\rho(T)$ and optical conductivity $\sigma_1(\omega)$.

While the quasiparticle concept has proven extremely powerful in describing good conductors, QPs become poorly defined in case of excessive scattering.
In metals, the scattering rate is expected to saturate when the mean free path approaches the lattice spacing, known as Ioffe-Regel-Mott (IRM) limit\cite{Gunnarsson2003,Hussey2004}. However, this bound is often exceeded  ($\rho\gg \rho_{IRM}$) in correlated, `Mott' systems\cite{Georges1996}.
In view of the apparent breakdown of Boltzmann transport theory, it is controversially discussed whether charge transport in such \textit{bad metals}\cite{Gunnarsson2003,Hussey2004} is in any way related to  QPs\cite{Deng2013,Milbradt2013,Deng2014}
 or whether entirely different excitations come into play\cite{Hartnoll2015}. By investigating the low-energy electrodynamics of a strongly correlated metal through comprehensive optical measurements, here we uncover the prominent role of QPs persisting into this anomalous transport regime,
providing
evidence for the former scenario.
Our results also demonstrate
the emergence of a `displaced Drude peak' (DDP, see inset of Fig.~\ref{scattering-sketch}) indicating incipient localization of QPs in a regime where their lifetime is already heavily reduced by strong electronic interactions.

\begin{figure}
\centering
\includegraphics[width=0.97\columnwidth]{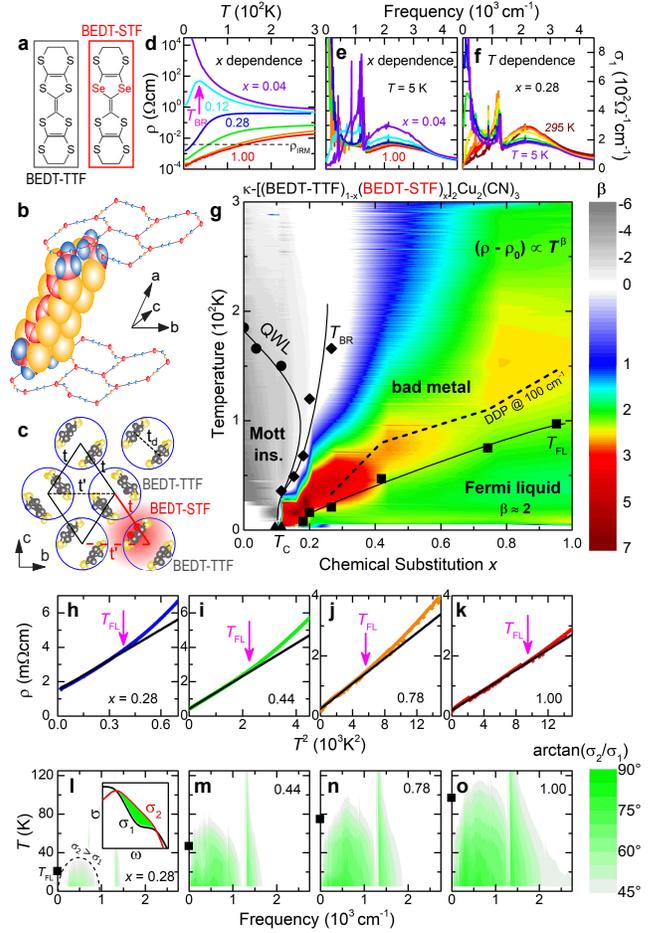}
\caption{\textbf{Mott transition to bad metal and Fermi liquid.}
\footnotesize{
\textbf{a-c,} Introducing selenium-containing BEDT-STF molecules (red) in the layered triangular-lattice compound \Cu\ locally increases the transfer integrals. \textbf{d,} This enhancement of electronic bandwidth by chemical substitution $x$ yields a textbook-like Mott MIT in the resistivity of \stf. \textbf{e,f,} The optical conductivity reveals the formation of a correlated metallic state when increasing $x$ and reducing $T$.  \textbf{g,}~Consistent with theory (see Fig. 3 of Ref. \cite{Vucicevic2013}), the resistivity exponent $\beta = {\rm d}(\ln\{\rho-\rho_0\})/{\rm d} (\ln\{T\})$ shows large negative values in the Mott-insulating state (grey) below the quantum Widom line (QWL). On the metallic side, Fermi-liquid like $\rho=\rho_0+ AT^2$ at low temperatures (green) crosses over to a bad metal above $T_{\rm FL}$, featuring a rapid increase beyond $\rho_{\rm IRM}$. $\beta \gg 2$ (orange-red) coincides with the displacement of the Drude peak from $\omega=0$ (DDP, dashed line). Metallic behavior is lost at $T_{\rm BR} $, where $\rho(T)$ has a maximum.
\textbf{h-k,} The $AT^2$ increase becomes steeper towards the 
MIT.
\textbf{l-o,} The phase angle determined from our optical experiments yields inductive semi-ellipses ($\arctan{\{\sigma_2/\sigma_1\}}>45^{\circ}$ corresponds to $\sigma_2>\sigma_1$) in $T-\omega$ domain as expected for a paradigmatic Fermi liquid\cite{Berthod2013,Tytarenko2015}. The sharp feature occuring for all compounds at 1200~\cm\ is a vibration mode\cite{SM}.
}
}
\label{structure-dc}
\end{figure}

\textbf{Results}  \quad
We have chosen the molecular charge-transfer salts \STF\ (abbreviated \stf), which constitute an ideal realization of the single-band Hubbard model on a half-filled triangular lattice.
In the parent compound  of the series ($x=0$), strong on-site Coulomb interaction $U\approx 2000$~\cm\ (broad maximum of $\sigma_1(\omega)$ in Fig.~\ref{structure-dc}e,f) gives rise to a genuine Mott-insulating state\cite{Pustogow2018,Dressel2018} with no magnetic order\cite{Shimizu2003} down to $T=0$. Partial substitution of the organic donors by Se-containing BEDT-STF molecules with more extended orbitals
\cite{Saito2018} increases the transfer integrals $t\propto W$ (Fig.~\ref{structure-dc}a-c).
As a result, the correlation strength $U/W$ is progressively reduced
with $x$, allowing us to tune the system through the "bandwidth-controlled" Mott metal-insulator transition (MIT), covering a wide range in $k_{\rm B}T/W$ and $\hbar\omega/W$ within the parameter ranges accessible in our transport and optical experiments\cite{SM}.

$\rho(T)$ of \stf\ (Fig.~\ref{structure-dc}d) reveals a textbook Mott MIT resembling the pressure evolution of \Cu,\cite{Kurosaki2005,Furukawa2015,Furukawa2018} which turns metallic around 1.3~kbar.
As $x$ rises further, Fermi-liquid behavior $\rho(T)=\rho_0 + AT^2$ stabilizes below a characteristic $T_{\rm FL}$ that increases progressively with $x$, while $A\propto (m^{\star})^2$ is reduced\cite{Jacko2009} as correlations diminish (Fig.~\ref{structure-dc}h-k). As common for half-filled Mott systems\cite{Vucicevic2013,Milbradt2013}, $\rho(T)$ rises faster than $T^2$ above $T_{\rm FL}$, seen by the effective temperature exponent $\beta\gg 2$ in Fig.~\ref{structure-dc}d, and exceeds $\rho_{\rm IRM}=hd/e^2\approx 4$~m$\Omega$\,cm ($h=2\pi\hbar$ is Planck's constant, $e$ the elementary charge and $d\approx 16$~\AA\ the inter-layer spacing). We note that the bad metal formed here does not exhibit a linear-in-$T$ resistivity that occurs in many other materials\cite{Hartnoll2015}.
Instead, metallic behavior is completely lost when temperature exceeds the kinetic energy of QPs at the Brinkman-Rice scale $k_{\rm B} T_{\rm BR}\simeq Z E_{\rm F}$,\cite{Brinkman1970,Deng2013} and $\rho(T)$ resembles a thermally activated semiconductor above the resistivity maximum\cite{Radonjic2012,Pustogow2019-percolation}.
Also in the optical conductivity we observe the transition from an insulator (${\rm d}\sigma_1/{\rm d}\omega>0$ at low frequencies) to a metal (${\rm d}\sigma_1/{\rm d}\omega\leq 0$), forming a QP peak at $\omega=0$ upon increasing $x$ and lowering $T$ (Fig.~\ref{structure-dc}e,f).
The phase diagram in Fig.~\ref{structure-dc}g summarizes the characteristic crossovers in \stf\ (black symbols), also including the quantum Widom line (QWL)\cite{Terletska2011,Vucicevic2013} that separates the Mott insulator with a well-defined spectral gap from the incoherent semiconductor at elevated temperatures.

The low-energy response of Fermi liquids and the corresponding quadratic scaling laws are well explored theoretically\cite{Gurzhi1959,Maslov2012,Berthod2013,Maslov2017}.
A scattering rate $\gamma\propto \omega^2$ implies an inductive  response characterized by
$\sigma_2>\sigma_1$, where $\sigma_1$ and $\sigma_2$ are the real and imaginary part of the optical conductivity, respectively. This occurs in the so-called `thermal' regime\cite{Berthod2013}, $\omega>\gamma$, delimited by semi-elliptical regions in $T-\omega$ domain, as recently reported in Fe-based superconductors\cite{Tytarenko2015}. Our optical data on \stf\ indeed reveal inductive behavior, signaled by characteristic semi-ellipses with a phase angle $\arctan{(\sigma_2/\sigma_1)}>45^{\circ}$ (Fig.~\ref{structure-dc}l-o), and occurring at temperatures where $\rho\propto T^2$ is seen in dc transport (Fig.~\ref{structure-dc}h-k), i.e. at $T<T_{\rm FL}$ (black squares). Concerning the  $\omega^2/T^2$ scaling  in Eq.~\ref{FL-Gurzhi}, Fermi-liquid theory predicts a `Gurzhi parameter' $p=2$ for the optical scattering rate $\gamma(T,\omega)$,\cite{Gurzhi1959} a quantity that can be extracted from the optical conductivity via extended Drude analysis.
Experimentally, values both in the range $1 \leq p \leq 2$ \cite{Nagel2012,Mirzaei2013,Stricker2014,Tytarenko2015}  and $p\geq 2$  \cite{Yasin2011}
have been found for $\gamma(T,\omega)$ in few selected materials. While purely inelastic scattering among QPs yields $p=2$, deviations towards $p=1$ have been assigned to elastic scattering off quasi-static impurities (dopants, $f$-electrons), for instance \cite{Maslov2017}.
It remains an open question how the $T$ and $\omega$ dependences, and the value of $p$, develop as correlations advance towards the Mott MIT.

\begin{figure}
\centering
\includegraphics[width=1.0\columnwidth]{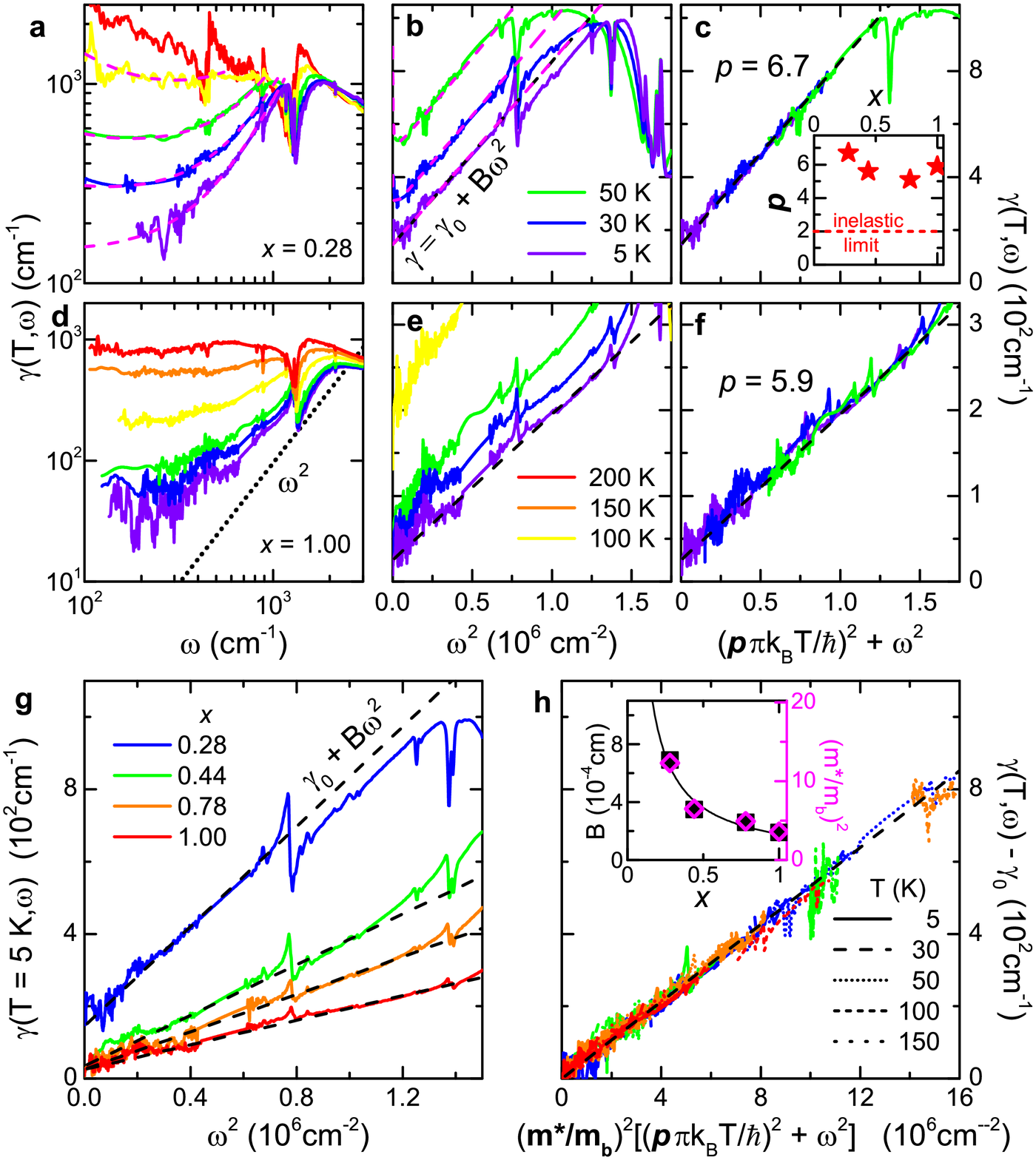}
\caption{\textbf{Fermi-liquid scaling of optical scattering rate obtained from extended Drude analysis.}
\footnotesize{
\textbf{a,d,} $\gamma(T,\omega)$ acquires a pronounced frequency dependence at low $T$, here shown for $x= 0.28$ and 1.00.  \textbf{b,e,} $\omega^2$ behavior persists well above $T_{\rm FL}$; note the quadratic frequency scales. Dashed pink lines in \textbf{a,b} are fits to Eq.~\ref{eq:DLC}. \textbf{c,f,} Curves recorded at different $T$ collapse on a generalized quadratic energy scale (see Eq.~\ref{FL-Gurzhi}) for a specific Gurzhi parameter $p>2$, as shown in the inset. \textbf{g,} Comparing $\gamma(\omega)$ at 5~K for $x\geq 0.28$ reveals that the slope of $B\omega^2$ increases towards the Mott MIT, similar to $AT^2$ in dc transport (Fig.~\ref{structure-dc}h-k). \textbf{h,} Rescaling the energy dependence by $( {m^{\star}}/{m_{b}})^2$ \cite{SM} collapses all data on a universal scaling curve, which follows from $B\propto( {m^{\star}}/{m_{b}})^2$. The 5, 30 and 50~K data are shown for all four substitutions (color code like in \textbf{g}); the scaling holds even for $T\geq 100$~K for $x\geq 0.44$. In panels \textbf{b,c,e,f,h} only the range below the vibrational features at 880 or 1200 \cm\ is considered.
}
}
\label{gamma}
\end{figure}

In \stf, the broadband response follows $\gamma\propto \omega^2$ at low $T$ (Fig.~\ref{gamma}), as expected from Eq.~\ref{FL-Gurzhi}, in all the compounds of the series ($x=$0.28, 0.44, 0.78 and 1.00) that also show Fermi-liquid behavior in $\rho(T)$.
The pronounced dip visible in the spectra around 1200~\cm\ stems from a vibration mode
with Fano-like shape in $\sigma_1(\omega)$ and does not affect the relevant low-frequency behavior (see Fig.~\ref{structure-dc}e,f and Ref. \cite{SM}).
Analogue to the increase of the slope $A$ in Fig.~\ref{structure-dc}h-k,  the $\omega^2$ variation of the scattering rate also becomes steeper as correlations gain strength (Fig.~\ref{gamma}g), i.e. the coefficient $B$ increases as $x$ is reduced. In both cases, the quadratic energy dependence (and any ${\rm d}\gamma/{\rm d}\omega>0$) appears only below $\gamma_{\rm IRM}\approx 1000$~\cm.

The stringent prediction Eq.~(\ref{FL-Gurzhi}) can be directly verified by adding the $T^2$- and $\omega^2$-dependences of $\gamma(T,\omega)$ to a common energy scale.
In Fig.~\ref{gamma}c,f, the curves at different $T$ do fall on top of each other upon  scaling via a Gurzhi parameter $p=6\pm 1$ for all \stf\ (inset of panel c)\cite{SM}.  Even more striking, multiplying the energy scale by $\left( {m^{\star}}/{m_{b}} \right)^2$ collapses the data of all four substitutions on one universal line (Fig.~\ref{gamma}h). This  manifestation of the Kadowaki-Woods relation\cite{Jacko2009}, $B\propto (m^{\star})^2$ (see inset), rules out any relevance of spinons near the Mott MIT\cite{Senthil2008,Furukawa2018}, in accord with dynamical mean-field theory (DMFT) results\cite{Lee2016}.
All in all, the observed scaling provides compelling evidence for the applicability of Landau's Fermi-liquid concept, in agreement with previous studies on unconventional superconductors\cite{Mirzaei2013,Stricker2014,Tytarenko2015}. The Gurzhi parameter significantly exceeds the inelastic limit ($p=2$),\cite{Maslov2017} indicating quasi-elastic backscattering processes (see Eqs.~\ref{eq:DLC} and~\ref{delta-gamma} below).

\begin{figure}[t]
\centering
\includegraphics[width=0.97\columnwidth]{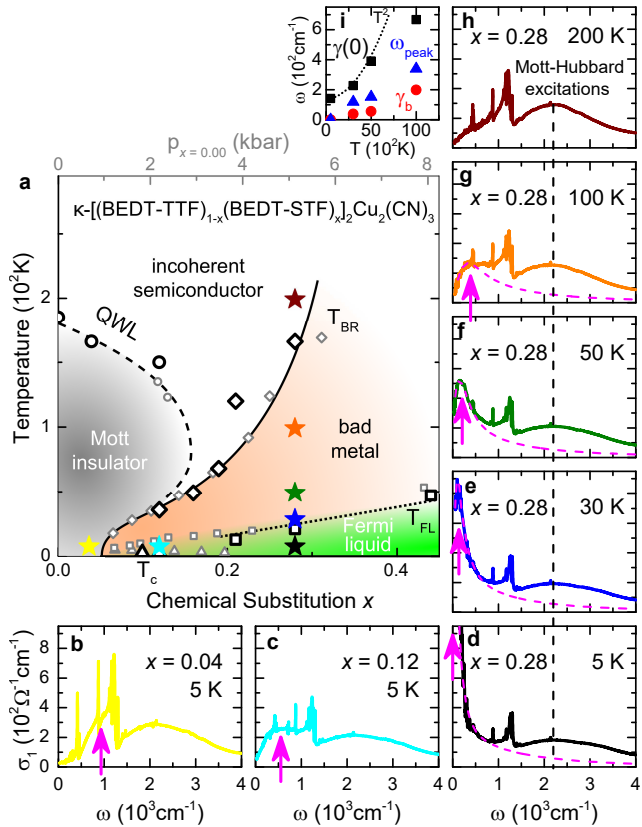}
\caption{\textbf{Displaced Drude peak linked to bad metal.}
\footnotesize{
\textbf{a,} Chemical substitution $x$ (black symbols) and physical pressure $p$ (grey, see Ref.~\cite{SM}) have the same effect on \Cu.
\textbf{b-h,} Evolution of $\sigma_1(T,\omega,x)$ through the phase diagram; stars with respective color indicate the position in \textbf{a}. Entering the bad-metallic phase for $T>T_{\rm FL}$ shifts the Drude peak away from $\omega=0$. The maximum broadens and hardens with $T$, until it dissolves at $T_{\rm BR}$. Approaching the insulator ($x=0.28\rightarrow 0.04$) at low $T$, the QP feature transforms into finite-frequency metallic fluctuations within the Mott gap\cite{Pustogow2018}.
\textbf{i,} Fit parameters (Eq.~\ref{eq:DLC}) for dashed pink lines in \textbf{d-g} and Fig.~\ref{gamma}a,b. Dotted line indicates $\gamma(0)=\gamma_0+B(p\pi k_{\rm B} T/\hbar)^2$, with $B$ and $p$ from Fig.~\ref{gamma}.
}
}
\label{sigma1-maxima}
\end{figure}

Having analyzed the QP properties and their dependence on electronic correlations,
we now want to evaluate how they behave when scattering increases as we
cross over from the Fermi liquid into a bad metal.
Fig. \ref{sigma1-maxima} displays $\sigma_1(\omega)$ at distinct positions in the $T$-$x$ phase diagram (stars in panel a); note the similarity between \stf\ (black symbols) and \Cu\ subject to pressure tuning (grey). For all $x\geq 0.28$ and $T<T_{\rm FL}$, the optical spectra feature a Drude-like peak centered at $\omega=0$, representing the QP response,
together with
a broad absorption centered at $U\approx 2000$~\cm\  originating from
electronic transitions between the Hubbard bands\cite{Pustogow2018},
as shown in Fig. \ref{sigma1-maxima}d (see also Fig.~\ref{structure-dc}e,f).
While the high-energy features show only weak dependence on $x$ and $T$,\cite{SM} a marked shift of spectral weight takes place within the low-frequency region.
The fingerprints of mobile carriers evolve upon moving away from the Fermi-liquid regime by either changing $x$ (Fig. \ref{sigma1-maxima}b-d) or increasing $T$ (d-h), until they
completely disappear both
in the Mott insulator (panel b) and in the incoherent semiconductor (panel h, $T>T_{\rm BR}=166$~K at $x=0.28$).

Closer scrutiny reveals that this gradual evolution of the low-frequency absorption is accompanied by the appearance of a dip at $\omega=0$, which occurs at $T\geq T_{\rm FL}$; this is also where the resistivity  becomes anomalous, deviating from $\rho\propto T^2$. The QP response then evolves
into a finite-frequency peak, that steadily shifts to higher $\omega$ and broadens with increasing $T$/reducing $x$  (arrows in Fig. \ref{sigma1-maxima}e-g and triangles in panel i). Such a displaced Drude peak eventually dissolves into the Hubbard band at  $T\sim T_{\rm BR}$.
From contour plots\cite{SM} of $\sigma_1(T,\omega)$ we can estimate the $T$-$\omega$ trajectory of the DDP above $T_{\rm FL}$ for the different substitutions: a peak frequency around 100~\cm\ (dashed line in Fig.~\ref{structure-dc}g) coincides with the steepest increase of the resistivity, i.e. the largest values of the exponent $\beta>2$.

The emergence and fading of the DDP at $T_{FL}$ and $T_{\rm BR}$, respectively, indicate that the observed behavior is tightly linked to the bad-metal response in the resistivity,  tracking the changes experienced by the QPs as the Fermi liquid degrades.
This physical picture is reminiscent of the recently introduced concept\cite{Deng2013,Milbradt2013,Deng2014} of `resilient' QPs, which persist beyond the nominal Fermi-liquid regime, but with modified (e.g. $T$-dependent) QP parameters.
Note that the DDP phenomenon observed here, that is not predicted by current theoretical descriptions of Mott systems\cite{Deng2013,Radonjic2010},
also impacts charge transport itself: for example, the values of $\sigma_1$ and $\gamma$ seen at finite frequency in our optical experiments, which yield correlation strengths $U/W\approx 1.3$ for $x\geq 0.28$,\cite{SM} are compatible with those computed by DMFT\cite{Radonjic2010}, but the measured dc resistivity increases way beyond the theory values --- a natural consequence of the drop of $\sigma_1$ at low frequencies upon DDP formation.

Building on the considerations above, we now show that our experimental observations
can be explained by an incipient localization of the carriers in the bad metal,  caused
by non-local, coherent backscattering corrections to semi-classical transport\cite{Smith2001,Fratini2016}.
We note that
related ideas have been
invoked to explain the bad metallicity and DDP observed in liquid metals\cite{Smith2001} and various correlated systems, including organics\cite{Takenaka2005}, cuprates\cite{Takenaka2003,Hwang2007} and other oxides\cite{Kostic1998}, but no systematic quantitative investigations have been provided to date.


In order to describe the experimental observations, we now introduce a model that assigns the modifications of the Drude peak to backscattering processes\cite{Smith2001,Fratini2016,Fratini2014,Fratini2020}:
\begin{equation}
\sigma(\omega)=\frac{\epsilon_0\omega_p^2}{\gamma-\gamma_b}
\left\lbrack
\frac{\gamma}{\gamma-i\omega}-\frac{\gamma_b}{\gamma_b-i\omega}
\right\rbrack .
\label{eq:DLC}
\end{equation}
Here the first term between brackets represents the standard metallic response
with the energy-dependent $\gamma$ from Eq. \ref{FL-Gurzhi}, where
$\omega_p$
is the plasma frequency.
The second term represents the leading finite-frequency correction beyond semiclassical transport, caused by additional  elastic or quasi-elastic processes.  Its  sign is opposite to that of the semiclassical Drude response, leading to a dip-peak structure in $\sigma_1(\omega)$ as illustrated in the inset of Fig. \ref{scattering-sketch}. The resulting peak frequency, $\omega_{peak}\simeq \sqrt{\gamma(0)\gamma_b}$,
gives a direct measure of the  backscattering rate $\gamma_b$.
Physically, the  "localization" corrections embodied in Eq.~\ref{eq:DLC} represent \textit{non-local}
interference processes, which can be viewed as finite-frequency precursors of a disorder-induced bound-state formation.

We have used Eq.~\ref{eq:DLC} to fit the finite-frequency spectra at the substitution $x=0.28$, where the DDP is most clearly identified in experiment.
The Fermi-liquid response has been extracted from Fig. \ref{gamma}, setting a constant $B=6.7\times 10^{-4}$ cm at all temperatures up to $T=100$~K. Importantly,
$\omega_p^2$ is also kept constant, compatible with the fact that the spectral weight associated with the QPs is conserved from the Fermi liquid to the bad-metallic region. The fits accurately describe the experimental data, as demonstrated by magenta lines for $\sigma_1(T,\omega)$ in Fig.~\ref{sigma1-maxima}d-g and for $\gamma(T,\omega)$ in Fig.~\ref{gamma}a,b.
Similar to the direct determination from the extended Drude analysis, the extracted  $\gamma(0)$ shows an initial $T^2$ dependence which is lost at $T\geq T_{\rm FL}$, as illustrated in Fig.~\ref{sigma1-maxima}i.  The parameter $\gamma_b \ll \gamma(0)$  shows a similar trend.

To get further microscopic insight,
 we isolate explicitly the anomalous scattering contributions
by writing
\begin{equation}
\delta\gamma(\omega,T)=\gamma(\omega,T)- \gamma_{\rm FL,2}(\omega,T),
\label{delta-gamma}
\end{equation}
where $\gamma(\omega,T)$ is the measured scattering rate, which has the general form Eq. (\ref{FL-Gurzhi}), and $\gamma_{\rm FL,2}$ is the strict Fermi-liquid prediction, i.e. Eq. (\ref{FL-Gurzhi}) with $p=2$. Direct comparison with  Eq. (\ref{FL-Gurzhi}) yields  $\delta\gamma(\omega,T)=B(p^2-2^2)(\pi k_BT/\hbar)^2$,  from which the following conclusions can be drawn.
First, the fact that we find a frequency-independent correction directly confirms the assumed (quasi)static nature of the anomalous scatterers.
Second, the fact that $p=6\pm 1$ is almost constant for all substitutions (Fig. \ref{gamma}c, inset) means that the strength of $\delta\gamma$ (in particular its variation with $x$) is  governed  by the QP scale embodied in the parameter $B$, i.e. $\delta\gamma\sim (\frac{m^{\star}}{m_{b}})^2 \sim  Z^{-2}$.
This observation stresses the key role of strong correlation effects in the vicinity of the Mott point.
Third, $p\gg 2$  implies that the anomalous contribution $\delta \gamma$ is dominant over the inelastic term, which consistently ensures that the corresponding localization effects are robust against the dephasing effects originating from inelastic QP scattering: whenever observed, the peak frequency $\omega_{peak}$ is much larger than the calculated dephasing term, $\sim B\omega^2$.


\textbf{Discussion}  \quad
The \STF\ series studied here realizes a continuous tuning through the genuine Mott MIT near $T\rightarrow 0$ that was previously not accessible by experiments applying physical pressure.
Our systematic investigation
of the electron liquid from the weakly interacting limit to the Mott insulator establishes Landau's QPs as the relevant low-energy excitations throughout the metallic phase. While demonstrating the
universality of Landau's QP picture,
the foregoing analysis also reveals an enhanced elastic scattering channel that fundamentally alters the QP properties in these materials. This is best visible within the bad-metallic regime, where it conspires with electronic correlations
in causing a progressive shift of the Drude peak to finite frequencies, indicative of
dynamical localization of the QPs. Our analysis also suggests that the same elastic processes may already set in  within the Fermi-liquid regime, causing deviations from the predicted $\omega^2/T^2$ scaling laws of QP relaxation.
These conclusions are largely based on a straightforward analysis of experimental data by a general theoretical model describing the optical response of charge carriers in the presence of incipient localization.
We now discuss possible scenarios to elucidate the possible microscopic origins. The key feature that requires explanation is the pronounced elastic scattering near the Mott point.


One
firmly established example leading to DDP behavior and anomalously high resistivities is the "transient localization" phenomenon found in crystalline organic semiconductors\cite{Fratini2020a}. There, soft lattice fluctuations provide a strong source of quasi-elastic randomness at room temperature,
causing coherent backscattering at low frequency and DDPs\cite{Fratini2020,Schubert2005}.
In the present \stf\ compounds the Debye temperature for the relevant inter-molecular phonons, $T_D \sim 30K$, is similar to that of organic semiconductors, compatible with transient localization at high $T$. However, this picture is difficult to reconcile with the observed DDP at very low $T$ close to the MIT that exhibits strong substitution dependence, indicating instead a clear connection with the Mott phenomenon.
Similar caveats would apply if lattice fluctuations were replaced by other soft bosons unrelated to the Mott MIT, such as charge/magnetic collective modes. While the latter can also give rise to finite-frequency absorption peaks, our clear assignment of the DDP to metallic QP rules out such a situation in the present case\cite{Caprara2002,Delacretaz2017}. 

An alternative possibility, that could reconcile the different experimental observations, is the physical picture of weakly disordered Fermi liquids\cite{Aguiar2005,Andrade2009}, motivated by the unavoidable structural disorder that accompanies chemical substitution\cite{Saito2018}. Although a complete theory for such a situation is still not available, existing studies\cite{Andrade2009} show that disorder directly affects the Fermi liquid, making its coherence scale $T_{\rm BR}$ spatially inhomogeneous with a broad distribution of local QP weights. In this case, one expects local regions with low $T_{\rm FL}$ to `drop out'  from the Fermi liquid and thus act essentially as vacancies --- dramatically increasing the elastic scattering as temperature is raised. While providing a plausible physical picture for $p>2$, this scenario would also be consistent with the observed scaling of $\gamma$ with $(\frac{m^{\star}}{m_{b}})^2$ upon approaching the Mott point, reflecting the gradual build-up of correlations in the disordered Fermi liquid.

Finally, we argue that long-range Coulomb interactions, that are usually neglected in theoretical treatments of correlated electron systems, could actually play a key role both in the present compounds as well as in other bad metals where DDPs have been reported\cite{Delacretaz2017}. The ability of non-local interactions in providing an effective disordered medium for lattice electrons has been recognized recently\cite{Pramudya2011,Mahmoudian2015,Rademaker2016,Driscoll2020},
with direct consequences on bad-metallic behavior\cite{Mousatov2019}.
The additional scattering channel associated with long-range potentials could well be amplified at the approach of the Mott transition,  due to both reduced screening and collective slowing down of the resulting randomness, possibly causing DDP behavior as observed here.


Since the gradual demise of quasiparticles is a general phenomenon in poor conductors,
displaced Drude peaks likely occur in many of them\cite{Delacretaz2017}.
In light of the present experiments, studying the interplay between electronic correlations and (self-induced) randomness appears to be a very promising route for understanding how good metals turn \textit{bad}.

\begin{center} -------------- \end{center}

\textbf{Methods}
Plate-like single crystals of \STF\ were grown electrochemically\cite{Saito2018} with a typical size of
$1 \times 1 \times 0.3$~mm$^{3}$; here BEDT-TTF stands for bis\-(ethylene\-di\-thio)\-tetra\-thia\-fulvalene and
BEDT-STF denotes
the partial substitution by selenium according to Fig~\ref{structure-dc}a.
The composition of $0\leq x\leq 1$ was determined by energy-dispersive x-ray spectroscopy\cite{Saito2018}.
The dc resistivity was recorded by standard four-point measurements; superconductivity was probed by magnetic susceptibility studies of polycrystalline samples using a commercial SQUID
and magnetoresistance measurements on single crystals. We performed complementary pressure-dependent transport experiments of the parent compound ($x=0$) providing the grey data points in Fig.~\ref{sigma1-maxima}a, see Ref.~\cite{SM}.
Since the compounds are isostructural they retain the highly-frustrated triangular lattice and do not exhibit magnetic order down to lowest temperatures; a hallmark of the spin-liquid state.
Using Fourier-transform infrared spectroscopy the optical reflectivity
at normal incidence was measured in the frequency range from 50 to 20000~\cm\
from $T=5$~K up to room temperature; here also the visible and ultraviolet regimes were covered up to 47\,600~\cm\ by a Woollam ellipsometer. The complex optical conductivity $\hat{\sigma}(\omega)=\sigma_1(\omega)+{\rm i}\sigma_2(\omega)$ is obtained via the Kramers-Kronig relations using standard extrapolations. Since the optical properties of both crystal axes provide similar information, we focus on the spectra acquired for
the polarization along the crystallographic $c$-axis.

The frequency-dependent scattering rate and effective mass are calculated via the extended Drude model\cite{Allen1977,Dressel2002}
\begin{equation}
\gamma (\omega) = \epsilon_0 \omega_{p}^2 \, {\rm Re}\left\{ [\hat{\sigma}(\omega)]^{-1} \right\}
\label{extended-Drude-gamma}
\end{equation}
\begin{equation}
\frac{m^{\star}(\omega)}{m_{b}}  = \frac{\epsilon_0 \omega_{p}^2}{\omega} \,  {\rm Im}\left\{-[\hat{\sigma}(\omega)]^{-1} \right\} \quad ,
\label{extended-Drude-mass}
\end{equation}
where $\omega_{p}=\sqrt{{Ne^2}/{\epsilon_0 m_{b}}}$  is the plasma frequency, comprising the charge-carrier density $N$ and band mass $m_{b}$; $\epsilon_0$ is the permittivity of vacuum and $e$ the elementary charge. $\omega_{p}$ is determined from the maximum of the dielectric loss function around 4000~\cm\ (see Ref.~\cite{SM}).

\textbf{Acknowledgments} We acknowledge fruitful discussions with D. L. Maslov, A. V. Chubukov, A. Georges and E. van Heumen. The project was supported by the Deutsche Forschungsgemeinschaft (DFG) via the projects DR228/39-3, DR228/41-1, DR228/48-1 and DR228/52-1. A.P. acknowledges support by the Alexander von Humboldt-Foundation through the Feodor Lynen Fellowship.

\textbf{Author Contributions}  A.P. and M.D. guided and conceived the experimental work. Optical experiments were conducted by A.P. and M.S.A., supported by Y.S.. dc transport measurements were performed by Y.S. and A.L.. Crystals were grown by Y.S. and A.K.. Circumstantial analysis of all results was carried out by A.P., with support from S.F. and in exchange with V.D. and M.D.. Theoretical work was performed by S.F., with contributions from V.D.. A.P., V.D., M.D. and S.F. discussed the data, interpreted the results and wrote the paper with input from all authors.

\textbf{Competing Interests} The authors declare that they have no competing interests.

\textbf{Data Availability} The authors declare that the data supporting the findings of this study are available within the paper and its Supplementary Information. Further information can be provided by A.P., M.D. or S.F..

\textbf{Correspondence} Correspondence and requests for materials should be addressed to A. Pustogow (andrej.pustogow@pi1.physik.uni-stuttgart.de), M. Dressel (dressel@pi1.physik.uni-stuttgart.de), or S. Fratini (simone.fratini@neel.cnrs.fr).



\pagebreak
\newpage

\setcounter{table}{0}
\setcounter{figure}{0}
\setcounter{equation}{0}
\renewcommand{\thefigure}{S\arabic{figure}}
\renewcommand{\thetable}{S\arabic{table}}
\renewcommand{\theequation}{S\arabic{equation}}

\onecolumngrid
\appendix

\section*{Supplementary Information}

\subsection*{Crystal Structure}

\begin{figure}[h]
	\centering
	\includegraphics[width=0.8\columnwidth]{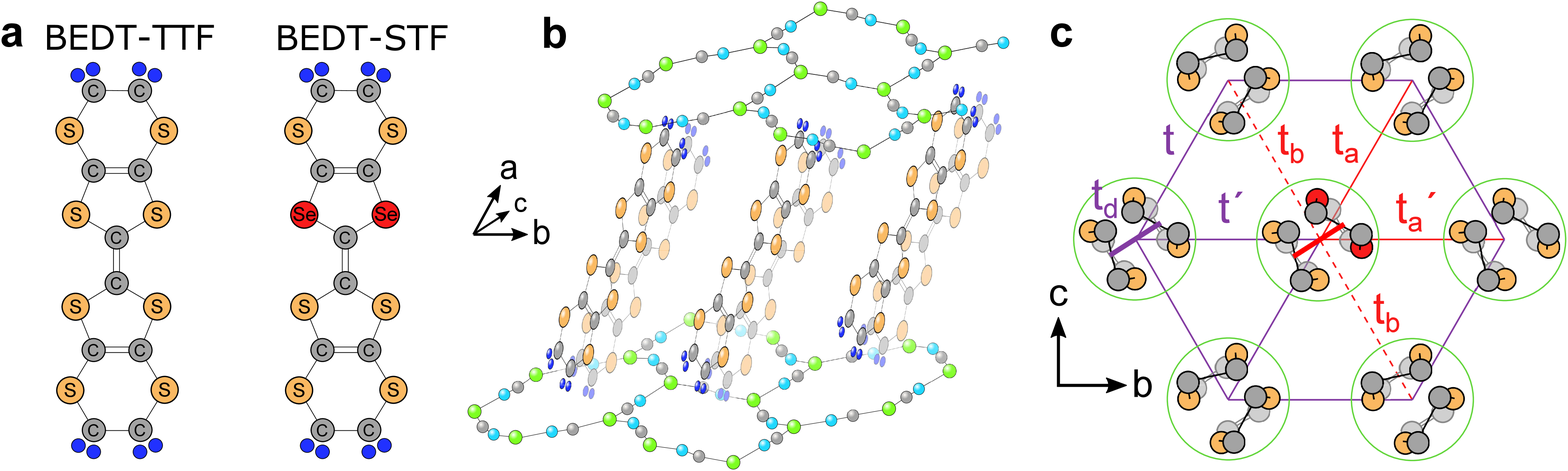}
	\caption{\textbf{Crystal structure. a,} Organic donor molecules BEDT-TTF and BEDT-STF. For the latter, two of the inner sulfur atoms are substituted by selenium. \textbf{b,} The crystal structure consists of layers of the donor molecules along the $bc$-plane that are separated by the polymeric Cu$_2$(CN)$_3$ anion network. \textbf{c,} Within the layers, the BEDT-TTF (BEDT-STF) dimers form a slightly anisotropic triangular arrangement with transfer integrals $t^{\prime}/t = 0.83$~\cite{Kandpal2009}, i.e.\ the lattice is subject to significant geometrical frustration. Substitution of BEDT-TTF by  BEDT-STF enlarges the transfer integrals (red) due to the extended orbitals of the selenium-containing BEDT-STF molecules.
Note, also the intra-dimer transfer integrals, $t_d\gg t$,$t^{\prime}$, are enhanced by substitution.}
	\label{fig:structure}
\end{figure}

Single crystals of \STF\ with the stoichiometries $x=0$, 0.04, 0.1, 0.12, 0.16, 0.19, 0.21, 0.25, 0.28, 0.44, 0.78 and 1.00 were prepared by standard electrochemical oxidation \cite{Geiser1991}.
BEDT-TTF stands for bis\-(ethylene\-di\-thio)\-tetra\-thia\-fulvalene and
BEDT-STF for bis-(ethylene\-di\-thio)\-di\-selenium-di\-thia\-fulvalene.
While BEDT-TTF molecules were purchased from Sigma-Aldrich, the
synthesis of BEDT-STF and the crystal growth were carried out at Hokkaido University in Sapporo. The two types of molecules are displayed in Fig.~\ref{fig:structure}a. In order to create different substitutions in the alloy series, the amount of donor molecules was pre\-selected; the actual value $x$ was determined {\it a posteriori} by energy-dispersive x-ray spectroscopy for each batch:  we compared the intensity of S atoms to that of Se atoms, using \Cu\ as a reference. More information on crystal growth can be found in Ref.~\onlinecite{Saito2018}.

The BEDT-TTF (resp.\ BEDT-STF) molecules form dimers because the intra-dimer transfer integral $t_d$ is significantly larger than the inter-dimer wave function overlaps $t$ and $t^{\prime}$ (Fig.~\ref{fig:structure}c). Per dimer one electron is donated to the [Cu$_2$(CN)$_3]_{\infty}^-$ anion sheets that separate the donor layers as shown in Fig.~\ref{fig:structure}b.
Within the $bc$ plane the dimers constitute a slightly anisotropic triangular lattice that is close to ideal geometrical frustration, $t^{\prime}/t = 0.83$~\cite{Kandpal2009} as illustrated in Fig.~\ref{fig:structure}c.
The system is subject to strong electronic correlations as,
overall, each dimer carries one hole with spin 1/2: on-site Coulomb repulsion among the charges realizes a Mott insulator at half filling~\cite{Kanoda2011}. In the alloys
the BEDT-TTF molecules are randomly substituted by BEDT-STF, which mainly leads
to an increase of the electronic bandwidth. It may also give rise to disorder in the transfer integrals, similar to chemical substitution or doping in other materials.

\subsection*{dc Resistivity}

\begin{figure*}
\centering
\includegraphics[width=1.0\columnwidth]{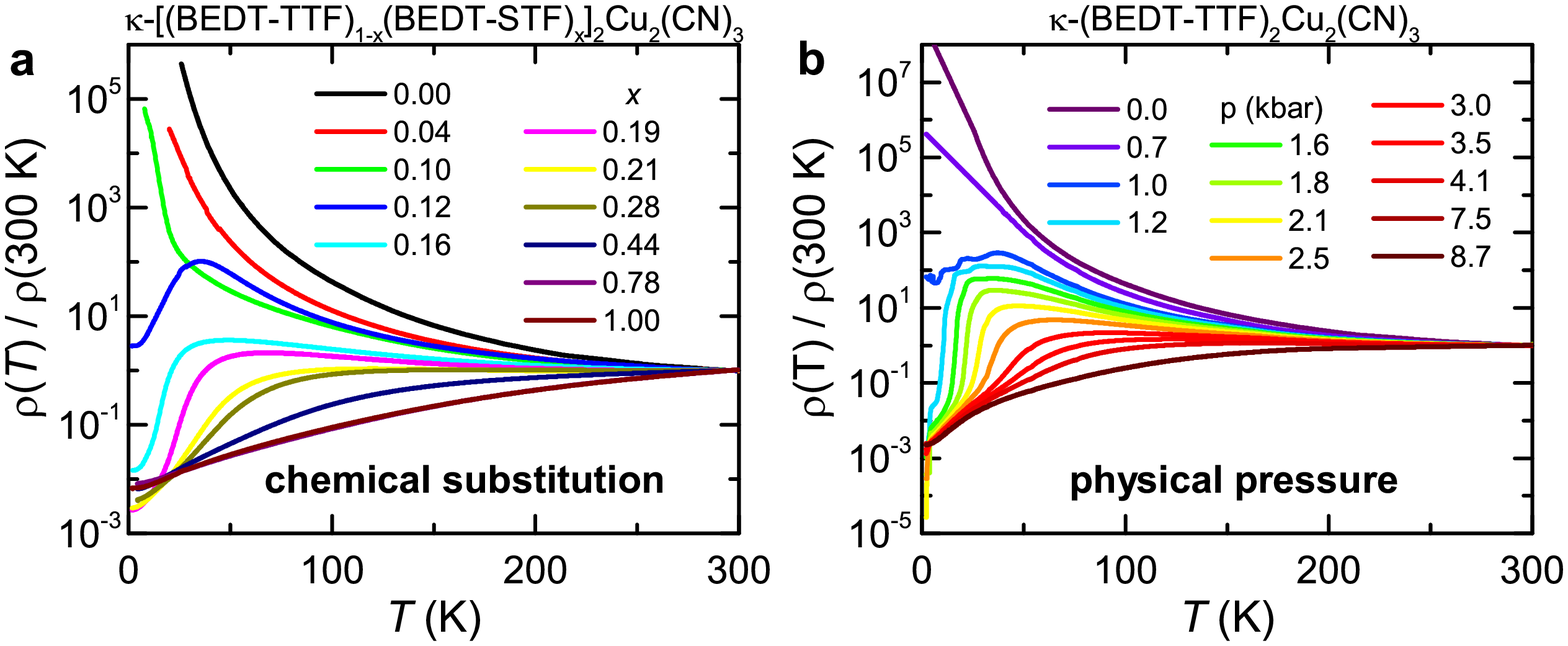}
\caption{\textbf{Substitutional and pressure evolution of the dc resistivity. a,}
The temperature dependence of the dc resistivity (normalized to $T=295$~K) is plotted
for the series \STF\ \cite{Pustogow2019-percolation}.
\textbf{b,} Temperature-dependent dc resistivity
was measured on \Cu\ for increasing hydrostatic pressure.
The resistivity $\rho(T)$ is successively reduced with higher pressure:
metallic behavior and superconductivity are observed at low temperatures,
in agreement with literature~\cite{Kurosaki2005,Furukawa2015,Furukawa2018}. The pressure values indicated here correspond to the lowest temperature.
During cooling the  pressure is reduced due to thermal contraction and
freezing of the pressure-transmitting medium (Daphne oil). {\it In-situ} calibration allows us to determine the actual pressure at each temperature, which is taken into account in Fig.~3a.
}
\label{SM_rho}
\end{figure*}
\begin{figure}[pbt]
\centering
\includegraphics[width=0.8\columnwidth]{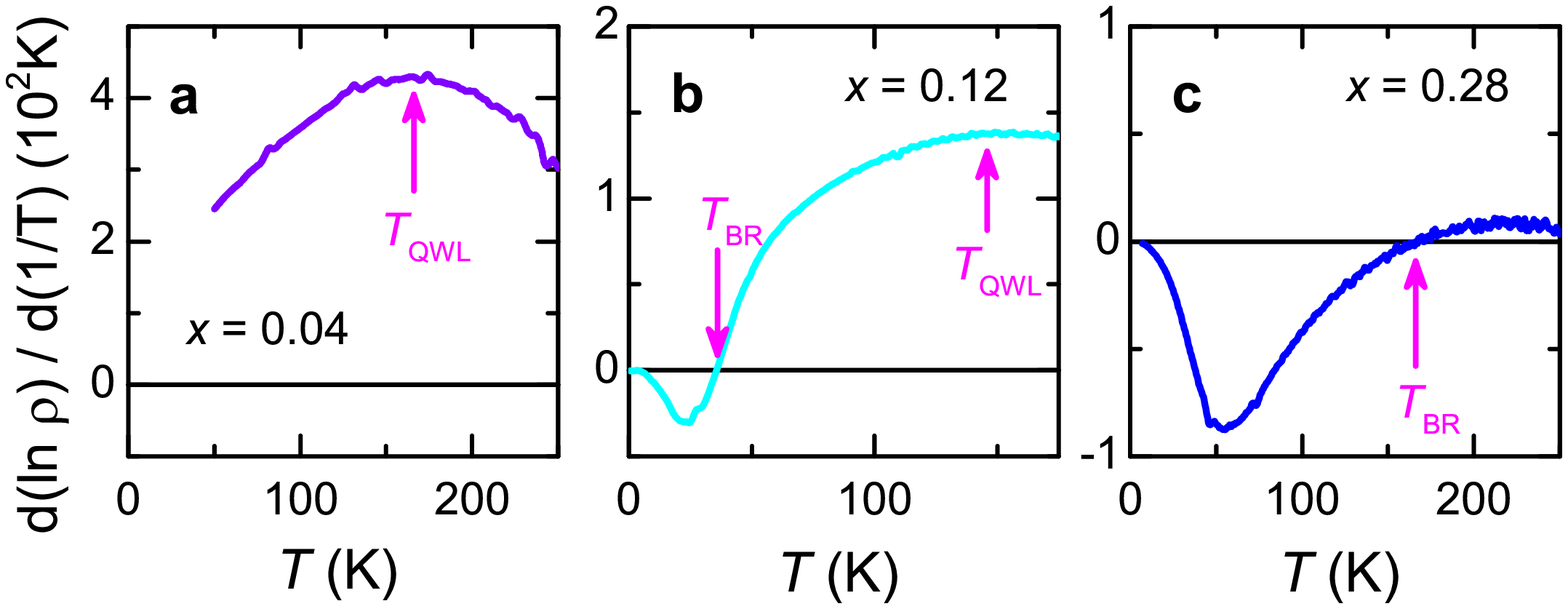}
\caption{\textbf{Transport gap defines quantum Widom line. a,b,}
For those compounds that show Mott-insulating behavior, $x=0.04$ and $x=0.12$, a maximum in the logarithmic derivative indicates the temperature when $\rho(T)$ crosses the quantum Widom line. \textbf{b,c,} The change of sign defines the Brinkman-Rice temperature $T_{\rm BR}$ as the onset of metallic transport ($d(\ln{\rho})/d(1/T)<0$).
}
\label{SM_transport-gap}
\end{figure}

Standard four-point dc transport experiments were carried out for the alloy series \STF\ in the temperature range from $T=300$~K down to 2~K. Fig.~\ref{SM_rho}a shows the complete set of resistivity curves $\rho(T)$ normalized to 300~K that were used to construct the phase diagrams in Fig.~2d and Fig.~4a. Prior to performing the logarithmic derivative, the residual resistivity $\rho_0$ in the limit $T\rightarrow 0$ has to be subtracted in order to yield the power-law exponent $\beta = {\rm d}(\ln\{\rho-\rho_0\})/{\rm d} (\ln\{T\})$.
While the Fermi-liquid temperature $T_{\rm FL}$ is reached
when the resistivity deviates from $\rho(T)=\rho_0+AT^2$,
the Brinkman-Rice temperature $T_{\rm BR}$ is defined when $\rho(T)$ goes through a maximum; both temperatures are indicated in Fig.~2.
In Fig.~\ref{SM_transport-gap}a,b we illustrate how the quantum Widom line is determined:
the steepest slope in the Arrhenius plot corresponds to the maximum in ${\rm d}(\ln\{\rho\})/{\rm d}(1/T)$; this is the distinctive criterion for $T_{\rm QWL}$ (cf.\ Ref.~\onlinecite{Pustogow2018}, Fig.~S2). As seen in Fig.~\ref{SM_rho}a,
a Mott-insulating behavior occurs only for $x\leq 0.12$.
The larger substitutions ($x\geq 0.12$) behave metallic below $T_{\rm BR}$;
this corresponds to a sign change in ${\rm d}(\ln\{\rho\})/{\rm d}(1/T)$ as
seen in Fig.~\ref{SM_transport-gap}b,c.
Furthermore for \STF\ with $x=0.10$ and 0.12 we find that around 2.5~-~3~K
superconductivity sets in at ambient pressure.

In addition to ambient-pressure dc transport experiments on the \STF\ alloy series,
we performed complementary pressure-dependent resistivity measurements on the parent compound ($x=0$); the results are displayed in Fig.~\ref{SM_rho}b.
Here the Mott insulator-metal transition occurs slightly above 1~kbar and is accompanied by superconductivity (onset at $T_{\rm c,max}\approx 5$~K), in agreement with previous work~\cite{Kurosaki2005,Furukawa2015,Furukawa2018}. The resulting values for $T_{\rm QWL}$, $T_{\rm FL}$, $T_{\rm BR}$ and $T_{\rm c}$ are included in Fig.~4a;
they line up well with the transition and crossover temperatures obtained via chemical substitution.



\subsection*{Optical Spectroscopy}

\begin{figure}
\centering
\includegraphics[width=0.7\columnwidth]{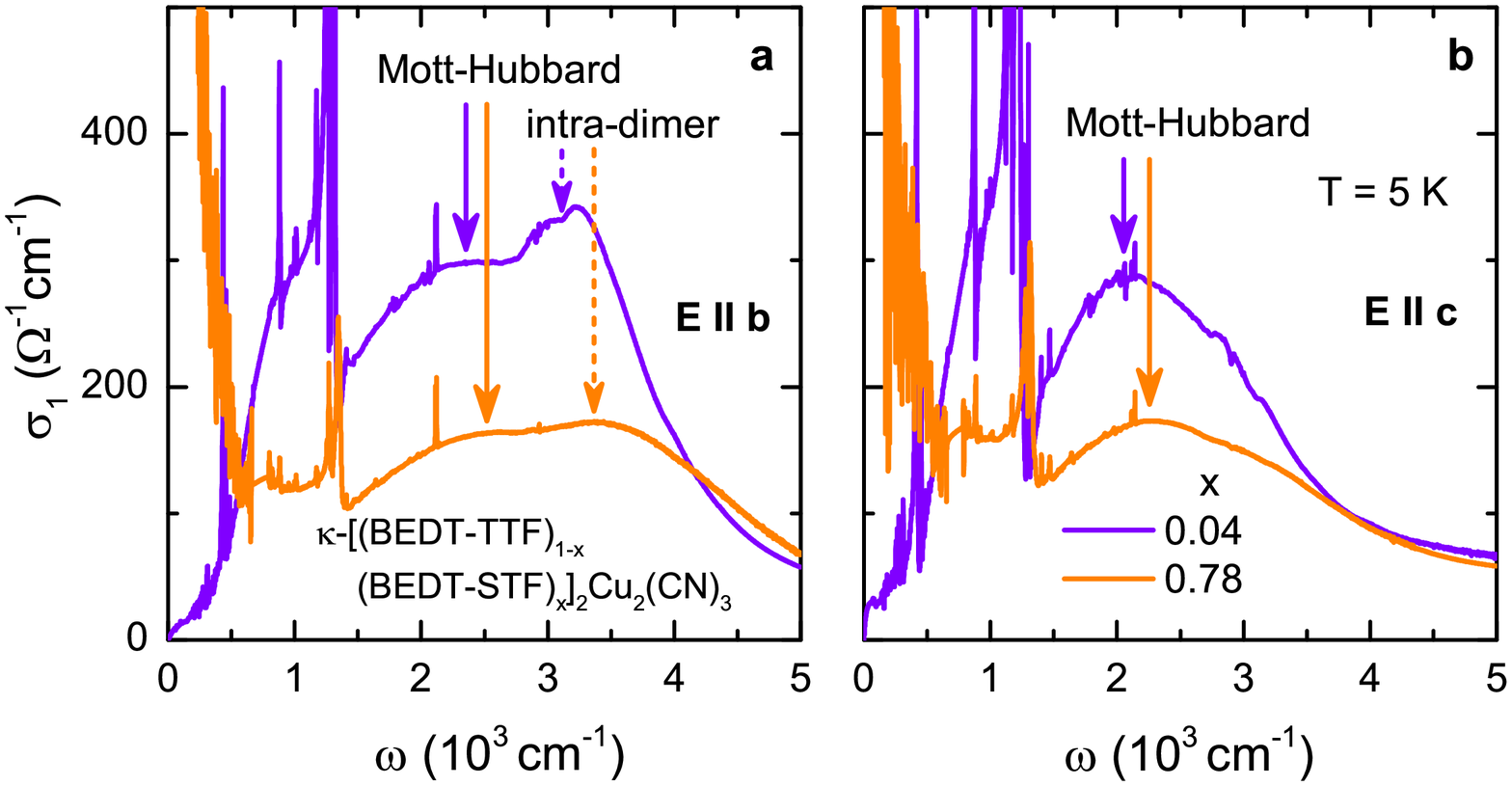}
\caption{\textbf{In-plane anisotropy.}
The anisotropy of the in-plane electrodynamic response of \STF\ between
(a) $E\parallel b$ and (b) $E\parallel c$
is exemplified for $x=0.04$ and 0.78 at $T=5$~K.
The main contribution along both crystal axes is due to Mott-Hubbard excitations between the dimers (solid arrows)~\cite{Faltermeier2007,Merino2008,Dumm2009,Ferber2014}. Overall, the Coulomb repulsion $U$, associated with the band maximum, and the corresponding electronic correlation strength $U/W$ are slightly larger for $E\parallel b$~\cite{Pustogow2018}. Local intra-dimer excitations show up above 3000~\cm\ (dashed lines) and are more pronounced along the $b$-direction as compared to $E\parallel c$. Both electronic contributions shift to higher energy with increasing STF content $x$.
}
\label{SM_bc-excitation-comparison}
\end{figure}

Optical spectroscopic experiments were performed on the substitution series \STF\ covering a frequency range of 50--20000~\cm\ and for temperatures from 5~K up to 295~K. Due to the small in-plane anisotropy of the transfer integrals (Fig.~\ref{fig:structure}c), the dc resistivity differs by a factor 2 between the $b$ and $c$ directions~\cite{Pinteric2014}. In Fig.~\ref{SM_bc-excitation-comparison} we compare the optical spectra of the two crystal axes. While for both polarizations the absorption exhibits a peak around 2500~\cm\ and 2000~\cm, respectively, along $b$-direction one can distinguish another feature above 3000~\cm\ which corresponds to intra-dimer excitations that do not contribute to charge transport~\cite{Faltermeier2007,Merino2008,Dumm2009,Ferber2014}. Apart from minor quantitative differences, metallic and insulating behavior expresses for both in-plane axes as exemplarily shown here for the substitutions $x=0.04$ and 0.78 in Fig.~\ref{SM_bc-excitation-comparison}. For the determination of $U$ and $W$ from the Mott-Hubbard transitions, we focused on the optical spectra acquired for $E\parallel c$.

In addition, a strong vibrational mode occurs around 1400~\cm\ related to intramolecular vibrations mainly involving C=C. Due to electron-molecular vibrational coupling, this feature
can become rather intense and acquire an asymmetric Fano shape, in particular for in the insulating regime \cite{Dressel2004}. Although these vibrational features might appear disturbing, we refrain from subtracting them off the spectra in an arbitrary fashion;
instead we present raw data whenever possible.

\begin{figure*}
\centering
\includegraphics[width=1\columnwidth]{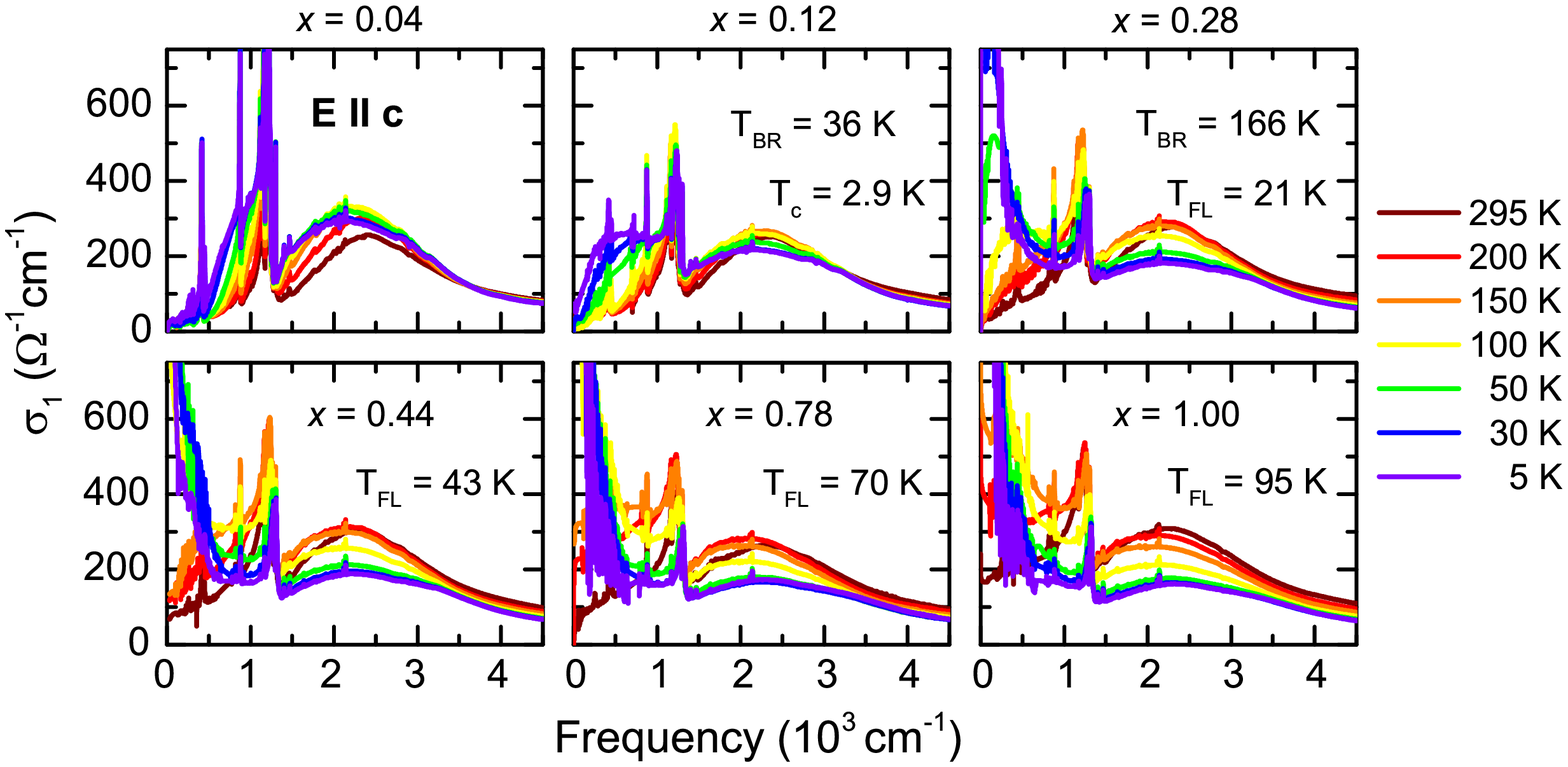}
\caption{\textbf{Temperature-dependent optical conductivity for all substitutions.}
The optical conductivity of \STF\ was determined in reflection experiments from room temperature down to $T=5$~K. As substitution $x=0.04,0.12,0.28,0.44,0.78,1.00$ increases, spectral weight successively accumulates at low frequencies coincident with the onset of metallic behavior below $T_{\rm BR}$. A Drude-like peak forms at zero frequency for $T<T_{\rm FL}$.
}
\label{SM_all-substitutions}
\end{figure*}

Fig.~\ref{SM_all-substitutions} displays the full set of optical data ($x=0.04,0.12,0.28,0.44,0.78,1.00$; $5~{\rm K} < T < 295$~K)
of the alloying series discussed in the present work. Similar to the parent compound~\cite{Kezsmarki2006,Elsasser2012}, the substitution $x=0.04$ exhibits an insulating-like optical conductivity with $\sigma_1(\omega\rightarrow 0)\rightarrow 0$ and ${\rm d}\sigma_1/{\rm d}\omega>0$ below the Hubbard transitions at all temperatures. Also for $x=0.12$ and 0.28, the high-temperature spectra ($T>T_{\rm BR}$) reveal insulating, or rather semiconducting properties -- at finite temperature there is always a non-zero offset to $\sigma_1$ at $\omega = 0$ due to thermal excitations across the gap. Below $T_{\rm BR}$, however, a significant amount of the spectral weight accumulates at low frequencies. Eventually, a Drude-like peak is established at $\omega=0$ for $T<T_{\rm FL}$ in all substitutions $x\geq 0.28$ that contain Fermi-liquid properties in dc transport.

\begin{figure}
\centering
\includegraphics[width=0.5\columnwidth]{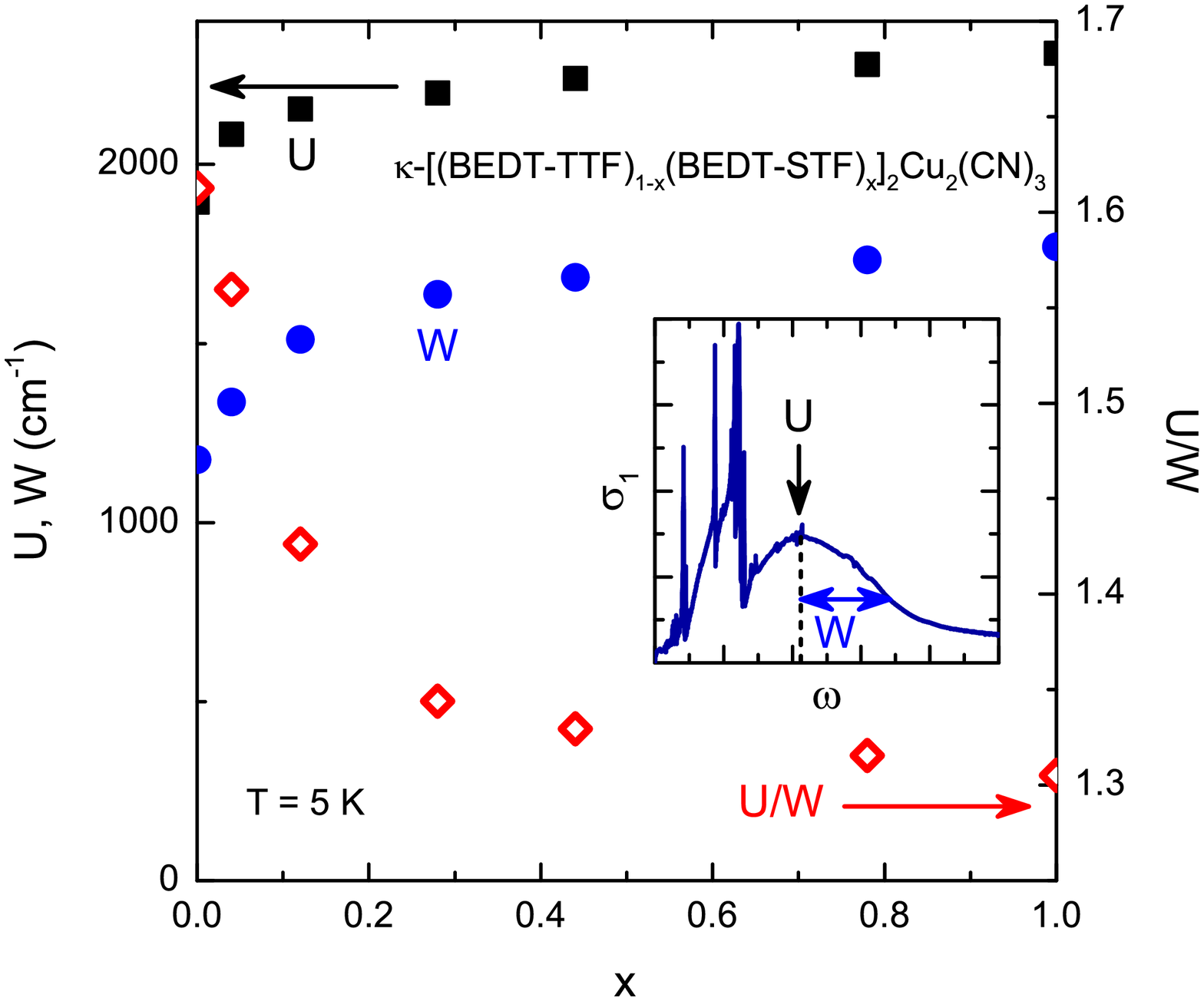}
\caption{\textbf{Band parameters $U$, $W$ and $U/W$ of \stf.}
The electronic bandwidth $W$ and on-site Coulomb repulsion $U$ were determined from the half-width at half maximum and the peak position, respectively. The increase of $W$ is more pronounced for $x \rightarrow 1$ leading to an overall reduction of the correlation strength $U/W$ (red symbols, right scale).
}
\label{SM_U-W}
\end{figure}

\begin{figure*}[t]
\centering
\includegraphics[width=1\columnwidth]{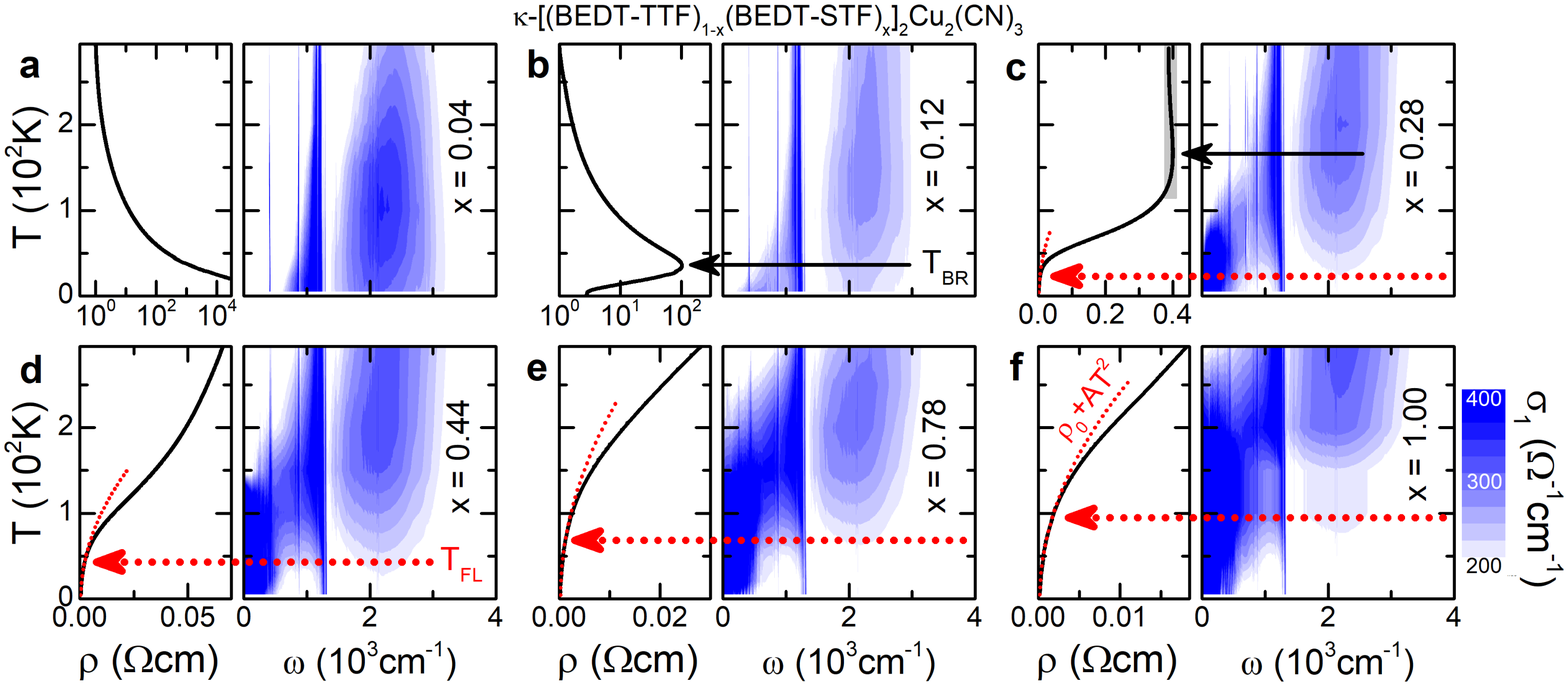}
\caption{\textbf{Temperature and frequency evolution of optical conductivity compared to dc resistivity in \stf.}
\textbf{a,} The insulating compound with $x=0.04$ exhibits an increase of the in-gap absorption upon cooling, despite monotonously increasing $\rho(T)$. Such non-thermal behavior of $\sigma_1(T,\omega)$ is also found in the parent compound ($x=0$) and assigned to metallic fluctuations in the insulating state close to the metal-insulator transition \cite{Pustogow2018}. \textbf{b,c,} Below $T_{\rm BR}$ a large part of the spectral weight separates from the Hubbard bands and gradually shifts to lower frequencies upon cooling, where it forms a metallic state at low temperatures. \textbf{c-f,} A well-defined quasiparticle peak forms at $T\leq T_{\rm FL}$. The $\rho\propto T^2$ behavior (red dotted lines) is unanimously linked to a Drude-like response at $\omega=0$, and the strong increase of resistivity in the bad metal arises from a successive spectral weight shift to finite frequencies upon warming. Metallic behavior is completely lost as the itinerant response merges with the Hubbard bands.
}
\label{SM_maxima-sigma-rho}
\end{figure*}

In Fig.~\ref{SM_U-W} we present the band parameters: the Coulomb repulsion $U$ and the electronic bandwidth $W$ are determined from the Mott-Hubbard transitions. As indicated in the inset, here we use the half-width at half-maximum on the high-frequency wing for a consistent determination of $W$ because the appearance of the metallic component for the higher substitutions conceals the low-frequency side. The enhancement of $W$ with increasing $x$ substantiates that chemical substitution primarily tunes the bandwidth via the larger transfer integrals of the BEDT-STF molecules.
In addition, also $U$ increases, which we assign to the accompanying enlargement of the intra-dimer transfer integrals, that is empirically used to approximate the on-site term via $U\approx 2t_d$~\cite{Kanoda1997,McKenzie1998}. Overall, the enhancement of $W$ is more pronounced than for $U$ leading to a sizeable reduction of $U/W$ with larger substitution, proving that the main effect of chemical substitution is tuning the bandwidth. This is in excellent agreement with the similarity of the Mott transition observed when applying physical pressure to \Cu\ \cite{Kurosaki2005,Furukawa2015,Furukawa2018}; the results of both experiments are compared in Fig.~4a.

In order to relate the optical response of \STF\ to their transport properties as a function of temperature, in Fig.~\ref{SM_maxima-sigma-rho} we compare $\rho(T)$ with a false-color contour plot of $\sigma_1(T,\omega)$ for each substitution. For $x\geq 0.28$ one can clearly see that the concentration of optical conductivity around $\omega=0$ coincides with the $T^2$-dependence of the resistivity (indicated by dotted red lines) in the Fermi-liquid state. Above $T_{\rm FL}$ the spectral weight associated with metallic quasiparticles gradually shifts to higher frequencies. Since around these temperatures the resistivity also exceeds the Ioffe-Regel-Mott (IRM) limit, the formation of a 'displaced Drude peak' (DDP) is clearly associated with the bad-metal phase. While a DDP was observed in a number of poor electrical conductors subject to pronounced electronic correlations~\cite{Delacretaz2017}, our results conclusively demonstrate that this phenomenon emerges directly from a Fermi liquid and is, thus, comprised of quasiparticles. This nourishes the idea of `resilient quasiparticles' discussed recently ~\cite{Deng2013,Deng2014}.

The spectral weight shift from low to high frequencies above $T_{\rm FL}$ approximately follows a linear temperature dependence. In Fig.~\ref{fig:structure} we estimated the temperature at which the DDP crosses $\omega=100$~\cm\ (dashed black line) by following the $T$-$\omega$ trajectory of the peak position in the contour plots of Fig.~\ref{SM_maxima-sigma-rho}c-f.
As the spectral contribution of itinerant carriers merges with the Hubbard excitations around 2000~\cm, the metallic properties get lost above $T_{\rm BR}$, as seen in Fig.~\ref{SM_maxima-sigma-rho}b,c for $x=0.12$ and 0.28. For the larger substitutions $x\geq 0.44$ this spectral feature broadens increasingly; and considerable weight remains at $\omega=0$ retaining metallic transport properties. Note, even well above $T=100$~K, the largest substitutions (Fig.~\ref{SM_maxima-sigma-rho}d-f) exhibit resistivities that deviate by less than a factor two from the low-temperature $\rho\propto T^2$ behavior. Naturally, this broad crossover from a Fermi liquid to a bad metal expresses similarly in $\sigma_1(\omega)$.

Finally, we point out that in Fig.~\ref{SM_maxima-sigma-rho}a-c a `metallic-like' increase of the low-frequency spectral weight is also observed for $T>T_{\rm BR}$ where the resistivity exhibits insulating properties, i.e.\ ${\rm d}\rho/{\rm d}T<0$ -- even seen for the completely insulating compound with $x=0.04$. In line with the discussion above, this contribution clearly originates from the metallic properties supporting our previous suggestion ~\cite{Pustogow2018} that the `non-thermal' increase of the low-energy spectral weight on the insulating side close to the Mott transition is associated to metallic fluctuations at finite frequencies. Overall, our optical results are in excellent agreement with dc transport, and provide information on the nature of metallic quasiparticles in the bad metal regime far beyond standard resistivity data.

\begin{figure*}[]
\centering
\includegraphics[width=1\columnwidth]{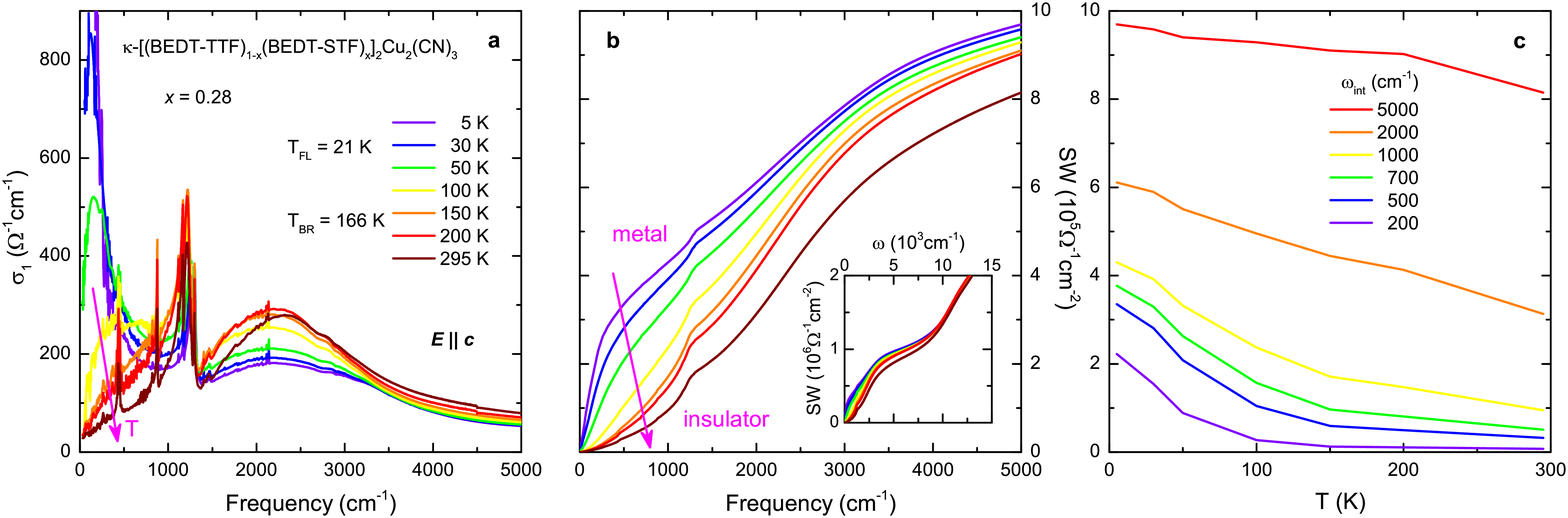}
\caption{\textbf{Spectral weight evolution for $x=0.28$. a,}
The optical conductivity of
$\kappa$-[(BEDT\--TTF)$_{0.72}$\-(BEDT\--STF)$_{0.28}$]$_2$\-Cu$_2$(CN)$_3$
is shown for all temperatures. \textbf{b,} The associated spectral weight ($SW$) clearly indicates the transition from a good metal at low temperatures to an insulating response above $T_{\rm BR}=166$~K; in other words spectral weight shifts from low to high frequencies. Inset: for all temperatures the $SW$ is conserved above $\omega\approx 10\,000$~\cm. \textbf{c,} Temperature dependence of the $SW$ integrated up to several distinct cut-off frequencies $\omega_{\rm int}$, as indicated.
}
\label{SM_SW-0.28-temperature}
\end{figure*}

In Fig.~\ref{SM_SW-0.28-temperature}a we exemplify the temperature dependence of the optical conductivity through the various conduction regimes for the example $x=0.28$. While the 5~K-spectrum comprises a Drude-like peak centered at $\omega=0$, at temperatures above $T_{\rm FL}=21$~K a DDP can be clearly identified; here seen for $T=30$, 50, and 100~K. As temperature increases, the spectral weight successively shifts to higher frequencies, as seen in the rising intensity around 2000--3000~\cm. The overall spectral weight, however, remains conserved for all temperatures around 10\,000~\cm, above which other types of excitations become dominant (Fig.~\ref{SM_SW-0.28-temperature}b inset). Panel c illustrates the temperature dependence of the spectral weight, when the integration is conducted from $\omega=0$ up to different cut-off frequencies $\omega_{int}$;
we identify a strong increase below $T_{\rm BR}=166$~K as metallicity sets in. Clearly, the changes are most pronounced at the lowest frequencies, in particular when $T_{\rm FL}$ is approached.

\subsection*{Extended Drude Analysis}

The extended Drude model~\cite{Allen1977,Dressel2002} assumes a frequency-dependent scattering rate $\gamma(\omega)$ and effective mass ${m^{\star}(\omega)}/{m_{b}}$. Both quantities are affected by electronic correlations via $Z<1$~\cite{Maslov2012,Berthod2013,Maslov2017}. The complex optical conductivity $\hat{\sigma}=\sigma_1+i\sigma_2$ of a correlated metal can be written as
\begin{align}
\hat{\sigma}(\omega)=\frac{\epsilon_0\omega_{p}^2}{\gamma}\frac{1}{1+i\omega (m^{\star}/m_{b})/\gamma},
\label{SM_extended-Drude}
\end{align}
where $\omega_{p}=\sqrt{{Ne^2}/{\epsilon_0 m_{b}}}$ is the plasma frequency, reflecting the charge-carrier density $N$ and band mass $m_{b}$; $\epsilon_0$ is the permittivity of vacuum and $e$ the elementary charge. Inversion of Eq.~\ref{SM_extended-Drude} yields the frequency-dependent scattering rate and effective mass
\begin{align}
\gamma (\omega) &= \epsilon_0 \omega_{p}^2 \, {\rm Re}\left\{ [\hat{\sigma}(\omega)]^{-1} \right\}
\label{SM_extended-Drude-gamma}\\
\frac{m^{\star}(\omega)}{m_{b}}  &= \frac{\epsilon_0 \omega_{p}^2}{\omega} \,  {\rm Im}\left\{-[\hat{\sigma}(\omega)]^{-1} \right\} \quad .
\label{SM_extended-Drude-mass}
\end{align}

\begin{figure}[t]
\centering
\includegraphics[width=0.5\columnwidth]{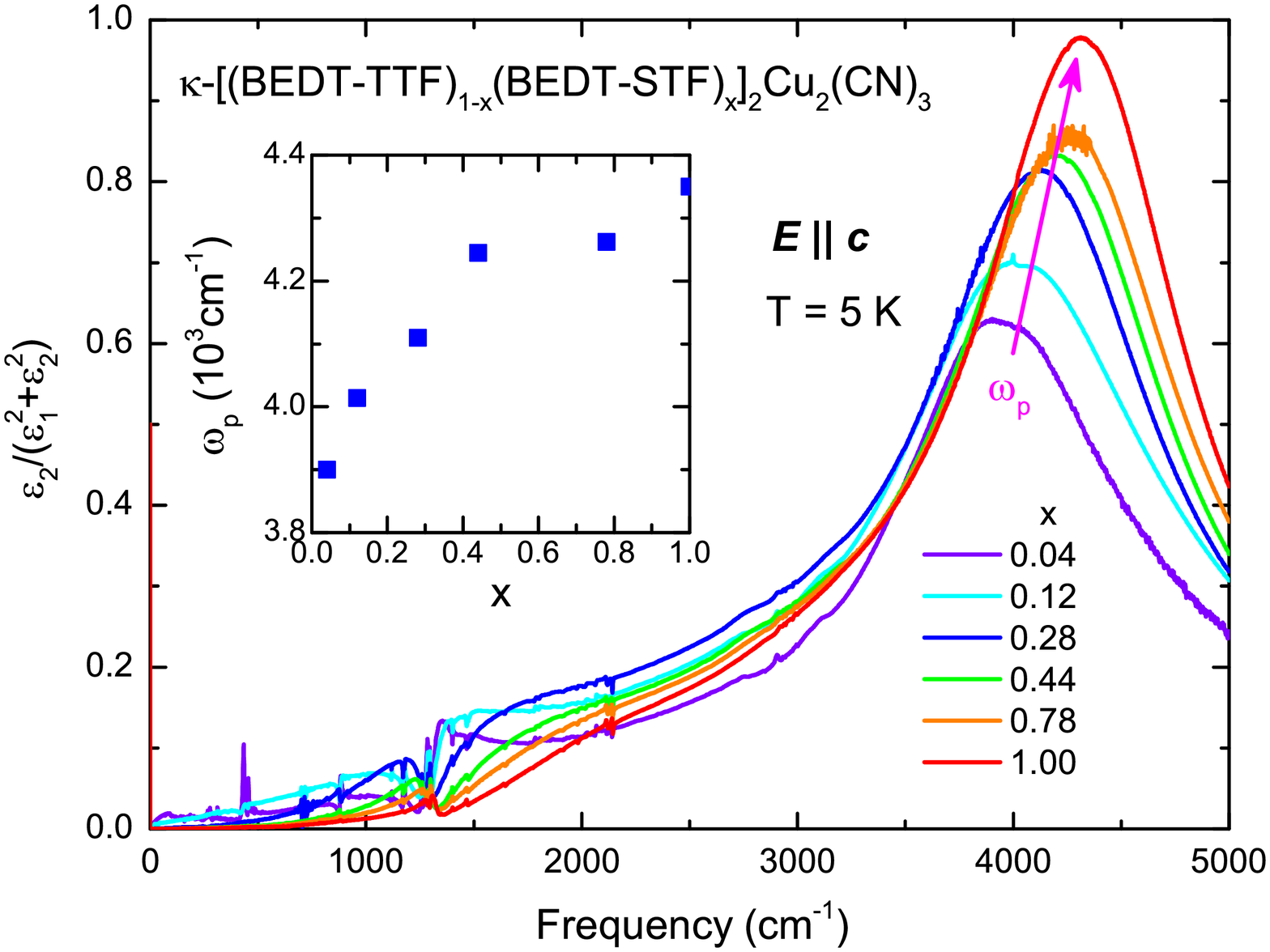}
\caption{\textbf{Dielectric loss function of \stf.}
The plasma frequency is determined from the peak in the dielectric loss function $-{\rm Im}\left\{\hat{\epsilon}^{-1}\right\}$ measured at $T = 5$~K. For all compounds \STF\ we obtain $\omega_{p}\approx 4000$~\cm. It slightly increases with substitution indicating a general reduction of the band mass $m_b$.
}
\label{SM_loss-function}
\end{figure}

The high-energy contributions are summarized as $\epsilon_{\infty}$,
which commonly is subtracted from the real part of the complex dielectric constant $\hat{\epsilon}=\epsilon_1+{\rm i}\epsilon_2$ before the extended-Drude analysis can be performed \cite{VanHeumen2007}. The particular values of $\epsilon_{\infty}$ range from 2.75 to 3.04 for $x=0.28$ - 1.00.
The plasma frequency $\omega_{p}$ of the conduction electrons in
Eqs.~(\ref{SM_extended-Drude-gamma}) and (\ref{SM_extended-Drude-mass}) is obtained from the peak of the dielectric loss function $-{\rm Im}\left\{\hat{\epsilon}^{-1}\right\} = \epsilon_2/(\epsilon_1^2+\epsilon_2^2)$, which occurs slightly above 4000~\cm\ for the compounds that exhibit metallic transport properties, displayed in Fig.~\ref{SM_loss-function}.


\begin{figure}[ptb]
\centering
\includegraphics[width=0.4\columnwidth]{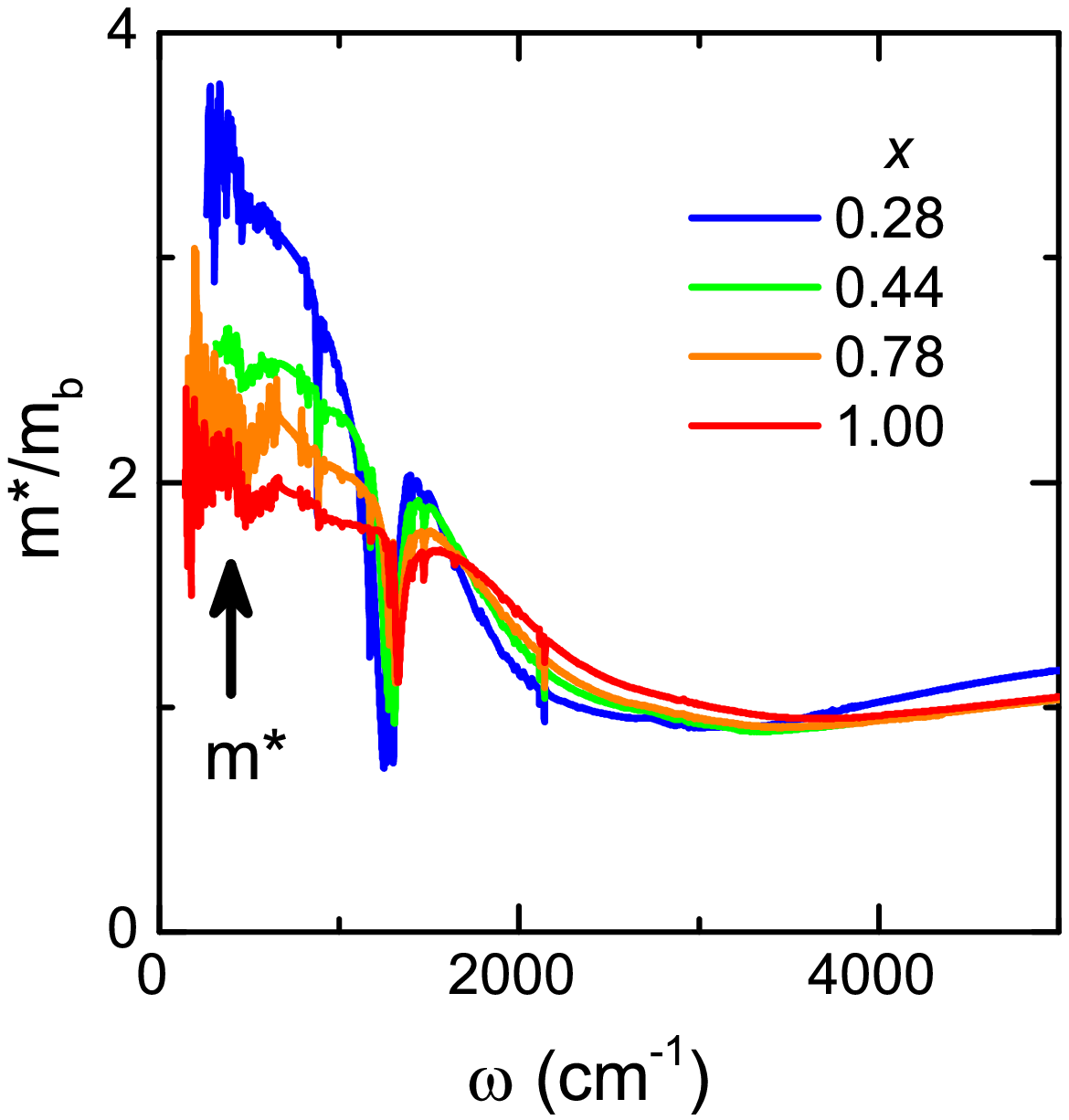}
\caption{\textbf{Frequency-dependent effective mass of \stf.}
The effective mass $m^{\star}/{m_{b}}$ of \STF\ is determined as a function of frequency by performing an extended Drude analysis. In all cases the low-temperature $m^{\star}(\omega)/m_{b}$
increases towards low frequencies where it forms a plateau that corresponds to the quasiparticle mass in the limit $\omega\rightarrow 0$.
}
\label{SM_effective-mass}
\end{figure}

In Fig.~\ref{SM_effective-mass}, we plot the effective mass normalized to the band mass, ${m^{\star}}/{m_{b}}$, determined at the lowest measured temperature ($T=5$~K) for the four compounds that exhibit Fermi-liquid properties, $x=0.28$, 0.44, 0.78 and 1.00.
Since $\omega_p$ shifts only by 5\%\ (Fig.~\ref{SM_loss-function}), $m_b$ is essentially identical for these four substitutions. The frequency dependence of $m^{\star}(\omega)$ is rather weak in the range 2000--4000~\cm, i.e. around the Mott-Hubbard transitions. Towards low frequencies, however, $m^{\star}(\omega)$ exhibits a pronounced increase
that eventually saturates at the static limit.
This plateau corresponds to the effective mass of long-lived quasiparticles.
Overall, our results reveal a frequency dependence of ${m^{\star}}/{m_{b}}$ similar to the optical data of the paradigmatic Fermi liquid Sr$_2$RuO$_4$ (cf.\ Ref.~\onlinecite{Stricker2014}, Fig.~SM5) and iron pnictides (cf.\ Ref.~\onlinecite{Tytarenko2015}, Fig.~3b); we regard the agreement as a strong confirmation of this concept~\cite{Nagel2012,Mirzaei2013,Tytarenko2015,Berthod2013,Maslov2012,Maslov2017}. As demonstrated in Fig.~3h, in the low-energy limit ($\omega\rightarrow 0$ and $T\rightarrow 0$) the effective mass enhancement follows the prefactor to the quadratic frequency dependence of the scattering rate, $\gamma\propto B\omega^2$, while approaching the Mott transition ($x=1.00\rightarrow 0.28$), yielding $B\propto\left({m^{\star}}/{m_{b}}\right)^2$ in full accord with the Kadowaki-Woods relation~\cite{Jacko2009}.

\begin{figure}[ptb]
\centering
\includegraphics[width=0.8\columnwidth]{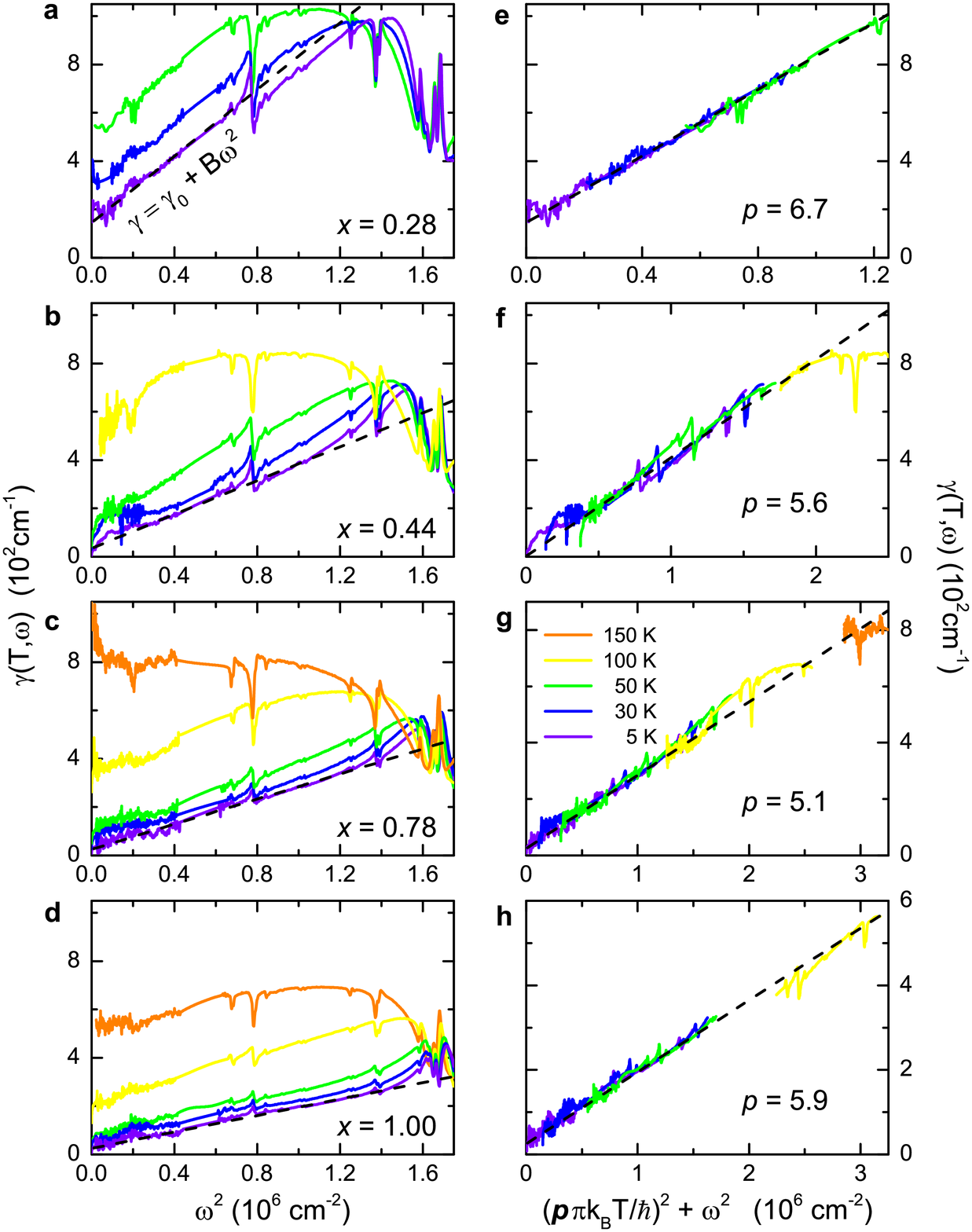}
\caption{\textbf{$\omega^2/T^2$ scaling of optical scattering rate. a-d,}
$\gamma(T,\omega)$ is displayed for the substitutions $x=0.28$, 0.44, 0.78, 1.00 on common axes. \textbf{e-h,} Renormalized to a generalized energy scale $(p \pi k_{\rm B}T/\hbar)^2+\omega^2$. Data are shown in the frequency range where Gurzhi scaling applies.
}
\label{SM_gamma-T-w_all}
\end{figure}

In Fig.~\ref{SM_gamma-T-w_all}a-d we display the optical scattering rate $\gamma(T,\omega)$ for the compounds ($x=0.28$, 0.44, 0.78, 1.00) that exhibit Fermi-liquid properties in their optical properties. Panels e-h display Gurzhi scaling on the generalized quadratic energy scale $(p \pi k_{\rm B}T/\hbar)^2+\omega^2$~\cite{Gurzhi1959,Maslov2012,Berthod2013,Maslov2017}. As indicated, we obtained a respective Gurzhi parameter $p=6\pm 1$ from our extended-Drude analysis.

\end{document}